\documentclass[
floatfix, 
pra, 
twocolumn, 
reprint, 
amsmath,amssymb, 
aps, 
]{revtex4}
\usepackage{graphicx}
\usepackage{dcolumn}
\usepackage{bm}
\usepackage{multirow}

\newcommand{\sig}{\sigma}
\newcommand{\lam}{\lambda}

\newcommand{\mr}{\mathrm}

\newcommand{\pri}{^\prime}

\newcommand{\FOR}{ \ \mathrm{for} \ }

\newcommand{\eco}{,\\}
\newcommand{\nnco}{\nonumber \\}
\newcommand{\peqn}[1]{
\begin{eqnarray}
#1
.\end{eqnarray}
}
\newcommand{\ceqn}[1]{
\begin{eqnarray}
#1
,\end{eqnarray}
}
\newcommand{\neqn}[1]{
\begin{eqnarray}
#1
\end{eqnarray}
}

\newcommand{\tbr}[1]{
\left\{ #1 \right\}
}
\newcommand{\rbr}[1]{
\left[ #1 \right]
}
\newcommand{\br}[1]{
\left( #1 \right)
}

\newcommand{\thmat}[9]{
\left(
\begin{array}{ccc}
#1 & #2 & #3 \\
#4 & #5 & #6 \\
#7 & #8 & #9
\end{array}
\right)
}

\usepackage{amscd}
\usepackage{multirow}
\usepackage{subfigure}
\usepackage{lscape}
\usepackage{color}

\usepackage{theorem}
\usepackage{makeidx}
\newtheorem{theo}{Theorem}
\newtheorem{defi}{Definition}
\newtheorem{lemm}{Lemma}
\newtheorem{coro}{Corollary}

\newcommand{\refE}[1]{Eq.~(\ref{eq:#1})}
\newcommand{\Eref}[1]{(\ref{eq:#1})}
\newcommand{\refErm}[1]{Eq.~$($\ref{eq:#1}$)$}

\newcommand{\refS}[1]{Sec.~\ref{sec:#1}}

\newcommand{\refA}[1]{Appendix~\ref{sec:#1}}
\newcommand{\Aref}[1]{\ref{sec:#1}}
\newcommand{\refF}[1]{Fig.~\ref{fig:#1}}
\newcommand{\Fref}[1]{\ref{fig:#1}}
\newcommand{\refT}[1]{Table~\ref{tb:#1}}
\newcommand{\refTH}[1]{Theorem~\ref{thm:#1}}
\newcommand{\refLM}[1]{Lemma~\ref{thm:#1}}
\newcommand{\refCO}[1]{Corollary~\ref{thm:#1}}

\newcommand{\labE}[1]{\label{eq:#1}}
\newcommand{\labS}[1]{\label{sec:#1}}
\newcommand{\labF}[1]{\label{fig:#1}}
\newcommand{\labT}[1]{\label{tb:#1}}
\newcommand{\labTH}[1]{\label{thm:#1}}

\newcommand{\dis}{\pi_0(H\cap G_0)}
\newcommand{\po}{\pi_1(G/H)}
\newcommand{\pog}{\pi_1(G)}
\newcommand{\poh}{\pi_1(H)}
\newcommand{\pt}{\pi_2(G/H)}
\newcommand{\ptg}{\pi_2(G)}

\newcommand{\pthgh}{\pi_3(G/H)}
\newcommand{\pthg}{\pi_3(G)}
\newcommand{\pthh}{\pi_3(H)}
\newcommand{\pim}{\pi_m(G/H)}

\newcommand{\KIM}{\mathrm{Ker} \ i_{\ast m-1}}
\newcommand{\CIM}{\mathrm{Coker} \ i_{\ast m}}
\newcommand{\COKo}{\mathrm{Coker} \ i_1^\ast}

\newcommand{\KER}{\mathrm{Ker} \ }
\newcommand{\KERb}[1]{\mathrm{Ker}\{#1\}}
\newcommand{\COK}{\mathrm{Coker} \ }
\newcommand{\COKb}[1]{\mathrm{Coker}\{#1\}}

\newcommand{\Ad}{\mathrm{Ad}}

\newcommand{\prf}{\times_f}
\newcommand{\SPAN}{\mr{Span}}

\newcommand{\Gh}{\mathfrak{h}}
\newcommand{\Gg}{\mathfrak{g}}
\newcommand{\uni}{\mathfrak{u}(1)}
\newcommand{\su}{\mathfrak{su}}
\newcommand{\so}{\mathfrak{so}}

\newcommand{\Bal}{\bm{\alpha}}

\newcommand{\mbZ}{\mathbb{Z}}

\newcommand{\mbR}{\mathbb{R}}

\newcommand{\Rs}{| R \rangle}
\newcommand{\Bs}{| B \rangle}
\newcommand{\Gs}{| G \rangle}

\begin{document}


\preprint{APS/123-QED}
\title{Influence of topological constraints and topological excitations: 
Decomposition formulas for calculating homotopy groups of symmetry-broken phases} 
\author{Sho Higashikawa${}^1$}
\author{Masahito Ueda${}^{1,2}$}
\affiliation{%
${}^1$Department of Physics, University of Tokyo, 7-3-1 Hongo, Bunkyo-ku, Tokyo 113-0033, Japan \\
${}^2$RIKEN Center for Emergent Matter Science (CEMS), Wako, Saitama 351-0198, Japan
}
\date{\today}

\newcommand{\te}{topological excitation}
\newcommand{\tinf}{topological influence}
\newcommand{\hg}{homotopy group}
\newcommand{\tcha}{topological charge}
\newcommand{\mon}{monopole}
\newcommand{\twsky}{two-dimensional skyrmion}
\newcommand{\thsky}{three-dimensional skyrmion}
\newcommand{\uint}{[0,1]}
\newcommand{\norm}[1]{\lVert \bm{#1} \rVert}
\newcommand{\Ex}[2]{\left\langle #1 \right\rangle_{#2}}
\newcommand{\change}[1]{\textcolor{red}{#1}}
\newcommand{\memo}[1]{\textcolor{blue}{\footnote{#1}}}

\begin{abstract}
A symmetry broken phase of a system with internal degrees of freedom 
often features a complex order parameter, 
which generates a rich variety of topological excitations 
and imposes topological constraints on their interaction (topological influence); 
yet the very complexity of the order parameter 
makes it difficult to treat topological excitations and topological influence systematically. 
To overcome this problem, 
we develop a general method to calculate homotopy groups and derive decomposition formulas 
which express homotopy groups of the order parameter manifold $G/H$ 
in terms of those of the symmetry $G$ of a system and those of the remaining symmetry $H$ of the state. 
By applying these formulas to general monopoles and three-dimensional skyrmions, 
we show that 
their textures are obtained through substitution of the corresponding $\mathfrak{su}(2)$-subalgebra for the $\mathfrak{su}(2)$-spin. 
We also show that a discrete symmetry of $H$ is necessary for the presence of topological influence 
and find topological influence on a skyrmion 
characterized by a non-Abelian permutation group of three elements 
in the ground state of an SU(3)-Heisenberg model. 
\end{abstract}
\pacs{Valid PACS appear here}
\maketitle


\section{\labS{introduction}Introduction}
Topological excitations create nontrivial spatial structures of the order parameter 
that cannot be removed by continuous deformation 
and are characterized by a topological charge. 
When a system has internal degrees of freedom 
as in spinor Bose-Einstein condensates (BECs) \cite{Kawaguchi12, Ueda10}, 
$p$-wave superfluids and superconductors \cite{Leggett75, Mackenzie03} 
and multiorbital electron systems \cite{Coqblin69, Kugel73, Tsunetsugu06, Lauchli06, Cazalilla14}, 
a symmetry broken phase has a complex order parameter, 
accommodating a rich variety of topological excitations. 
Examples include fractional and non-Abelian vortices \cite{Salomaa85, Semenoff07, Kleinert89}, 
skyrmions \cite{Skyrme61, Skyrme62}, 
Shankar skyrmions \cite{Shankar77, Volovik77a} and knot solitons \cite{Volovik77a, Faddeev97}. 
Several different types of topological excitations have been experimentally observed 
in condensed matter and ultracold atomic systems: 
skyrmions in chiral magnets \cite{Muhlbauer09, Yu10} and quantum Hall ferromagnets \cite{Barrett95, Schmeller95, Aifer96}, 
half-quantum vortices in $p$-wave superconductors \cite{Jang11} and liquid ${}^3$He \cite{Autti16}, 
and knot solitons in liquid crystals \cite{Chen13}. 
In particular, 
ultracold atomic gases offer an ideal playground for the study of topological excitations 
due to high controllability of experimental parameters; 
here the controlled generations of vortices \cite{Abo-Shaeer01, Seo15, Seo16}, 
skyrmions \cite{Leslie09, Choi12}, 
monopoles \cite{Ray14, Ray15}, 
and a knot soliton \cite{Hall16} have been demonstrated. 
Yet another remarkable feature arising from internal degrees of freedom is the coexistence of different types of topological excitations, 
which leads to non-conservation of individual topological charges due to \tinf \ \cite{Volovik77, Kleman77, Kobayashi12a}. 
For example, the A phase of superfluid ${}^3$He 
can simultaneously accommodate a half-quantum vortex and a monopole. 
When the latter makes a complete circuit of the former, 
the topological charge of the latter changes its sign \cite{Leonhardt00}.

In mathematical parlance, the set of topological charges constitutes a homotopy group of the order parameter manifold (OPM) $G/H$, 
where $G$ and $H$ are the symmetry of a system under consideration and the remaining symmetry of its state, respectively. 
Topological charges classify textures of an order parameter of topological excitations; 
two textures can continuously transform into each other 
if and only if their topological charges are the same. 
While the complexity of $G/H$ leads to the richness of \te s, 
it makes the calculation of homotopy groups involved, 
the understanding of textures highly nontrivial, 
and the analysis of \tinf \ difficult. 
While topological influence of a vortex on a topological excitation 
is known to be described by the action of the fundamental group $\po$ on the $m$th homotopy group $\pim$ \cite{Volovik77, Kleman77, Mermin79, Kobayashi12a}, 
where $m$ is the spatial dimension in which the texture of the topological excitation varies, 
general conditions for its presence are yet to be clarified. 
For topological influence on a monopole or a skyrmion, 
only one type is known, 
in which the \tinf \ changes the sign of the topological charge of a monopole and that of a skyrmion \cite{Volovik77, Mermin78, Leonhardt00, Kobayashi12a, Schwarz82b}. 

In the present paper, 
we develop a general method to calculate the homotopy group $\pim$ of the order parameter manifold $G/H$ 
by deriving a formula which expresses $\pim$ in terms of $\pi_m(G)$ and $\pi_m(H)$. 
Since the homotopy groups can be determined systematically for Lie groups \cite{Brocker85, Hall15}, 
$\pim$ and the corresponding textures can be determined through the formula. 
By applying the derived formulas for $m=2$ and $3$, 
we show that 
the texture of a general monopole and that of a general three-dimensional skyrmion 
are obtained from that of a monopole in a ferromagnet and those of a knot soliton or a Shankar skyrmion, respectively, 
through substitution of an appropriate $\su(2)$-subalgebra in $G$ for the $\su(2)$-spin. 
Consequently, their \tcha s are described by a set of integers distinguished by co-roots \cite{Brocker85, Hall15} 
which label different $\su(2)$-subalgebras in $G$. 

We also obtain the necessary and sufficient condition for the appearance of non-Abelian vortices 
and prove the absence of topological influence on a \thsky. 
We find that possible types of topological influence on a monopole or a skyrmion 
can be identified with the Weyl group \cite{Brocker85, Hall15} of $G$, 
where only one type is shown to be allowed if $G$ is U(1), SU(2), SO(3), or their direct product. 
Moreover, we find topological influence on skyrmions characterized by a non-Abelian permutation group of three elements in the ground state of an SU(3)-Heisenberg model \cite{Tsunetsugu06, Lauchli06, Toth10, Bauer12}, 
in which three types of skymions exchange their types through topological influence. 

This paper is organized as follows. 
In \refS{3}, 
we derive a decomposition formula for $\pim$ for an arbitrary dimension $m$. 
In \refS{4}, 
we derive simplified formulas for $\pim$ with $m = 1,2$, and $3$, 
and determine the texture of a general monopole and that of a general three-dimensional skyrmion. 
In \refS{5}, 
we analyze the conditions for the presence of topological influence. 
In \refS{6}, 
we discuss the non-Abelian topological influence on a skyrmion. 
In \refS{conclusion}, 
we conclude this paper. 
Some mathematical proofs are relegated to the appendices to avoid digressing from the main subject. 
\refA{proof1} proves a lemma on the third \hg \ of a compact Lie group used in \refS{3}. 
Appendices \Aref{proof2} and \Aref{proofcor2} prove formulas for $\pim$ and $\pt$, respectively, discussed in \refS{4}. 
\refA{prooftheo2} proves a theorem concerning \tinf \ on a general topological excitation discussed in \refS{5}. 
\refA{proof3} proves a corollary concerning \tinf \ on a \mon \ or a skyrmion discussed in \refS{5}. 

\section{\labS{3} Decomposition formula for homotopy groups of order parameter manifolds}

\subsection{\labS{3.A} Homotopy groups of a Lie group}
We first introduce the Cartan canonical form and the lattices of a compact Lie group, 
by means of which the first, second, and third homotopy groups are determined. 
When the parameter space of $G$ is (not) finite, $G$ is said to be (non-)compact.  
If $G$ includes translational symmetry, $G$ is non-compact. 
However, for the calculation of homotopy groups, 
$G$ can be replaced without loss of generality by its compact subgroup 
constituted from internal and rotational symmetries through the following isomorphism: 
\ceqn{
\pi_m(G) \simeq \pi_m(G_{\mr{int}} \times G_{\mr{rot}})
\FOR \forall m \ge 1 
\labE{reduction}
}
where $G_{\mr{int}}$ and $G_{\mr{rot}}$ are the internal and rotational symmetries of $G$, respectively, 
and $\simeq$ denotes the group isomorphism. 
The relation \Eref{reduction} can be proved as follows. 
The symmetry $G$ is, in general, constituted from 
an internal symmetry $G_{\mr{int}}$ and a space symmetry $G_{\mr{space}}$, 
i.e., $G  = G_{\mr{int}} \times G_{\mr{space}}$. 
The former is compact, while the latter may not be. 
Since $\pi_m\rbr{\mr{SO}(d,1)} \simeq \pi_m\rbr{\mr{E}(d)} \simeq \pi_m\rbr{\mr{SO}(d)}$ and $\pi_m(\mathbb{R}^d) \simeq 0$ for any $m \ (\ge 1)$ and any spatial dimension $d$ 
for the Lorentz group $\mr{SO}(d,1)$, the Euclid group $\mr{E}(d)$ and the translation group $\mathbb{R}^d$, 
the translational part of the symmetry does not contribute to homotopy groups. 
Then, we have 
\ceqn{
\pi_m(G) &\simeq& \pi_m(G_{\mr{int}}) \oplus \pi_m(G_{\mr{space}})
\nnco
 &\simeq& \pi_m(G_{\mr{int}}) \oplus \pi_m(G_{\mr{rot}})
\nnco
 &\simeq& \pi_m(G_{\mr{int}} \times G_{\mr{rot}})
}
where $\oplus$ denotes the direct sum. 
The first and third isomorphisms follow from the relation $\pi_m(X \times Y) \simeq \pi_m(X) \oplus \pi_m(Y)$.

\subsubsection{\labS{3.A.1} Cartan canonical form and lattices of a compact Lie group}
The Lie algebra $\Gg$ of a compact Lie group $G$ has a convenient basis called the Cartan canonical form \cite{Georgi99}: 
\ceqn{
\Gg = \{ \{ H_j \}_{j=1}^{r}, \{E_{\Bal}^R, E_{\Bal}^I \}_{\Bal \in R_+} \}
\labE{Cartanform}
}
where $E_{\Bal}^R := (E_{\Bal} + E_{\Bal}^\dagger)/\sqrt{2}$ and $E_{\Bal}^I := (E_{\Bal} - E_{\Bal}^\dagger)/(\sqrt{2} i)$ 
are the real and imaginary parts of the raising operator $E_{\Bal}$, 
and $r$ is the rank of $\Gg$. 
The Cartan canonical form \Eref{Cartanform} is 
a generalization of the basis of the $\su(2)$-Lie algebra $\{S_3, \{S_1, S_2\}\}$, 
and decomposes the generators of the Lie algebra into the off-diagonal matrices $\{E_{\Bal}^R, E_{\Bal}^I \}_{\Bal \in R_+}$ and the diagonal ones (Cartan generators) $\{ H_j \}_{j=1}^{r}$, 
where $\Bal$ is an $r$-dimensional real vector known as a positive root 
and $R_+$ denotes the entire set of positive roots. 
The positive roots are introduced to distinguish different $\su(2)$-subalgebras in $\Gg$. 
It is known that any positive root can be expressed as a linear combination of the $r$ positive roots known as simple roots, 
which we denote as $\tbr{\Bal_j}_{j=1}^r$. 
Two matrices $E_{\Bal}$ and $E_{-\Bal} = E_{\Bal}^\dagger$ are generalizations of 
the raising and lowering operators $S_+ := S_1 + i S_2$ and $S_- := S_1 - i S_2$ of the $\su(2)$-spin vector $\bm{S} = (S_1, S_2, S_3)$. 
Physically $\Bal$ describes the difference between two quantum numbers. 
When $E_{+\Bal} \ (E_{- \Bal})$ is applied to a state, its quantum number changes by $\Bal \ (- \Bal)$, 
as $S_+ \ (S_-)$ changes the magnetic quantum number of a spin state by $+1 \ (-1)$. 

Together with the Cartan generator $H_{\Bal}$ defined by 
$H_{\Bal} := \sum_{j=1}^r (\Bal)_j H_j$ 
with $\Bal = ((\Bal)_1, (\Bal)_2, \cdots, (\Bal)_r)^T \in \mbR^r$ ($T$ denotes the transpose of a vector), 
the two generators $E_{\Bal}^R$ and $E_{\Bal}^I$ satisfy the following commutation relations: 
\peqn{
&&\rbr{E_{\Bal}^R, E_{\Bal}^I} = i (\Bal, \Bal) H_{\Bal}, 
\nnco
&&\rbr{H_{\Bal}, E_{\Bal}^R} = i (\Bal, \Bal) E_{\Bal}^I, 
\nnco
&&\rbr{E_{\Bal}^I, H_{\Bal}} = i (\Bal, \Bal) E_{\Bal}^R
\labE{commrels}
}
We define the co-root $\Bal^c$ as a dual vector to each positive root $\Bal$ 
and the corresponding generator $H_{\Bal^c}$ as follows: 
\peqn{
&&\Bal^c :=  {2 \Bal \over (\Bal,\Bal)}
\labE{co-root}
\eco
&&H_{\Bal^c} := \sum_{j=1}^r (\Bal^c)_j H_j
}
One can see from \refE{commrels} that a triad $\bm{S}_{\Bal}$ defined by 
\neqn{
\bm{S}_{\Bal} := \br{S_{\Bal,1},S_{\Bal,2}, S_{\Bal,3}}
 := \br{ {E_{\Bal}^R \over (\Bal, \Bal)}, {E_{\Bal}^I \over (\Bal, \Bal)}, {H_{\Bal^c} \over 2} }
\labE{GM}
}
forms an $\su(2)$-subalgebra satisfying the following commutation relations: 
\ceqn{
\rbr{S_{\Bal,a},S_{\Bal,b}} = i \epsilon_{abc} S_{\Bal,c}
\FOR a,b,c = 1,2,3
\labE{comm}
}
where $\epsilon_{abc}$ is the three-dimensional Levi-Civita symbol which is a totally antisymmetric unit tensor of rank three. 
We refer to $\bm{S}_{\Bal}$ as a generalized $\su(2)$-spin vector 
by analogy with the ordinary $\su(2)$-spin vector $\bm{S}$ \cite{Higashikawa16a}. 
The integral lattice $L_G$ and the co-root lattice $L_G^c$ of $G$ are defined 
in terms of the Cartan generators of $\Gg$ and those of the co-roots as follows: 
\ceqn{
&&L_G := \tbr{H_{\bm{t}} \in \Gg | \exp(2 \pi i H_{\bm{t}}) = e}
\labE{integrallattice}
\eco
&&L_G^c := \tbr{\left. \sum_{\Bal} n_{\Bal} H_{\Bal^c} \in \Gg \right| n_{\Bal} \in \mbZ, \Bal \in R_+}
\labE{corootlattice}
}
where $H_{\bm{t}} \in \Gg$ for $\bm{t} \ (\in \mbR^r)$ is defined by 
$H_{\bm{t}} := \sum_{j=1}^r t_j H_j$ 
with $\bm{t} = (t_1, t_2, \cdots, t_r)^T$ ($T$ denotes the transpose of a vector). 
Both $L_G$ and $L_G^c$ form Abelian groups under the addition of matrices.

Consider an example of $\Gg = \su(3)$, 
which is generated by the following nine generators: 
\peqn{
&&S_{RG,1} = {1 \over 2} \thmat{0}{1}{0}{1}{0}{0}{0}{0}{0}, \ 
S_{RG,2} = {1 \over 2} \thmat{0}{-i}{0}{i}{0}{0}{0}{0}{0}, 
\nnco
&&S_{RG,3} = {1 \over 2} \thmat{1}{0}{0}{0}{-1}{0}{0}{0}{0}, \ 
S_{GB,1} = {1 \over 2} \thmat{0}{0}{0}{0}{0}{1}{0}{1}{0}, 
\nnco
&&S_{GB,2} = {1 \over 2} \thmat{0}{0}{0}{0}{0}{-i}{0}{i}{0}, \ 
S_{GB,3} = {1 \over 2} \thmat{0}{0}{0}{0}{1}{0}{0}{0}{-1}, 
\nnco
&&S_{BR,1} = {1 \over 2} \thmat{0}{0}{1}{0}{0}{0}{1}{0}{0}, \ 
S_{BR,2} = {1 \over 2} \thmat{0}{0}{i}{0}{0}{0}{-i}{0}{0}, 
\nnco
&&S_{BR,3} = {1 \over 2} \thmat{-1}{0}{0}{0}{0}{0}{0}{0}{1}
\labE{su3generators}
}
The corresponding Cartan canonical form is constituted from 
the following three generalized $\su(2)$-spin vectors 
\peqn{
&&\bm{S}_{RG} = (S_{RG,1}, S_{RG,2}, S_{RG,3}), 
\nnco
&&\bm{S}_{GB} = (S_{GB,1}, S_{GB,2}, S_{GB,3}), 
\nnco
&&\bm{S}_{BR} = (S_{BR,1}, S_{BR,2}, S_{BR,3})
} 
Note that the three diagonal generators $S_{RG,3}, S_{GB,3}$, and $S_{BR,3}$ 
are not linearly independent because $S_{RG,3} + S_{GB,3} + S_{BR,3} = 0$. 
Three root vectors $\Bal_{RG}, \Bal_{GB}$, and $\Bal_{BR}$ 
corresponding to generators $\bm{S}_{RG}, \bm{S}_{GB}$, and $\bm{S}_{BR}$ 
are given by 
\neqn{
\Bal_{RG} = (1,-1,0), \
\Bal_{GB} = (0,1,-1), \
\Bal_{BR} = (-1,0,1). 
\nnco
\labE{su3roots}
}
Since the lengths of these vectors are all $\sqrt{2}$, 
the corresponding co-roots $\Bal_{RG}^c, \Bal_{GB}^c$, and $\Bal_{BR}^c$ 
are given from \refE{co-root} by 
\peqn{
\Bal_{RG}^c = \Bal_{RG}, \
\Bal_{GB}^c = \Bal_{GB}, \
\Bal_{BR}^c = \Bal_{BR}
\labE{su3co-roots}
}
From direct calculations using Eqs.~\Eref{su3generators}, \Eref{su3roots}, and \Eref{su3co-roots}, 
one can show that the integral lattice $L_{\mr{SU}(3)}$ and the co-root lattice $L_{\mr{SU}(3)}^c$ coincide 
and that they are isomorphic to the triangular lattice [see \refF{corootlattice}]: 
\neqn{
&&L_{\mr{SU}(3)} = L_{\mr{SU}(3)}^c 
\nnco
&=& \tbr{\left. \sum_{a=\mr{RG,GB,BR}} m_a \Bal_a^c \right| m_a \in \mbZ, \sum_{a=\mr{RG,GB,BR}} \Bal_a^c = 0}
. \nnco
}

\begin{figure}
\centering
\includegraphics[width=6cm, bb=0 0 360 360]{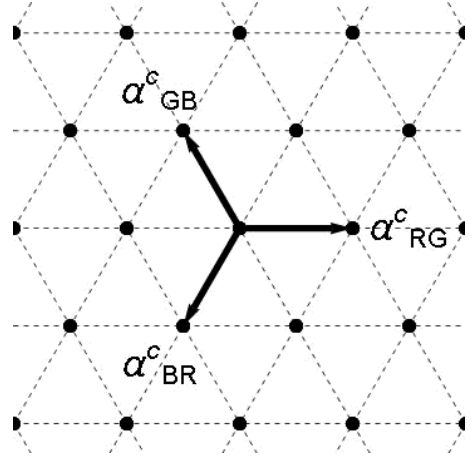}
\caption{\labF{corootlattice} 
Co-root lattice $L_G^c$ of $G = \mr{SU}(3)$ generated by the three co-roots $\Bal_{RG}^c, \Bal_{GB}^c$, and $\Bal_{BR}^c$ in \refE{su3co-roots}. 
Since $\Bal_{RG}^c + \Bal_{GB}^c + \Bal_{BR}^c = 0$, 
$L_G^c$ is isomorphic to the triangular lattice. 
We take the system of coordinates such that $\Bal_{RG}^c = (1,0), \Bal_{GB}^c = (- 1/2, \sqrt{3}/2)$, and $\Bal_{BR}^c = (- 1/2, - \sqrt{3}/2)$. 
}
\end{figure}

\subsubsection{\labS{3.A.2} First, second, and third homotopy groups of a compact Lie group}
It is known that $L_G^c$ is an Abelian subgroup of $L_G$ and 
that the quotient group $L_G/L_G^c$ is isomorphic to $\pog$ \cite{Brocker85, Hall15}: 
\peqn{
\pi_1(G) \simeq L_G/L_G^c
\labE{homG1}
}
While $L_G$ describes all loops on $G$, 
$L_G^c$ describes only those loops on $G$ that can continuously transform into a trivial one, 
so the quotient space naturally gives $\pog$. 
To be concrete, let us consider an element $H_{\bm{t}}$ of $L_G$ corresponding to a loop defined by 
\peqn{
g_{1,n}(\phi) := \exp(i \phi H_{\bm{t}})
\FOR \phi \in [0,2\pi]
\labE{homG1.1}
}
The map $g_{1,n}$ indeed describes a loop on $G$, 
since $g_{1,n}(0) = g_{1,n}(2\pi) = e$ from \refE{integrallattice}. 
The triviality of loops in $L_G^c$ can be checked by 
considering one of its generator $H_{\Bal^c}$ and the corresponding loop $g_{1,d}(\phi) := \exp\br{i \phi H_{\Bal^c}}$. 
This loop can continuously transform into a trivial one 
through $g_{\Bal}^{(2)}(\theta,\phi)$ defined by 
\ceqn{
g_{\Bal}^{(2)}(\theta,\phi) := \mr{e}^{ - i \theta S_{\Bal,2} }\mr{e}^{i \phi S_{\Bal,3}}\mr{e}^{i \theta S_{\Bal,2}}\mr{e}^{i \phi S_{\Bal,3}}
\labE{deform}
}
where $\theta \ (\in [0,\pi])$ is the parameter of the deformation. 
In fact, we have 
\ceqn{
g_{\Bal}^{(2)}(\theta = 0,\phi) &=& \mr{e}^{i \phi S_{\Bal,3}}\mr{e}^{i \phi S_{\Bal,3}} = g_{1,d}(\phi)
\labE{gBal1}
\eco
g_{\Bal}^{(2)}(\theta = \pi,\phi) &=& \mr{e}^{ - i \pi S_{\Bal,2} }\mr{e}^{i \phi S_{\Bal,3}}\mr{e}^{i \pi S_{\Bal,2}}\mr{e}^{i \phi S_{\Bal,3}}
\nnco
&=& \mr{e}^{- i \phi S_{\Bal,3}} \mr{e}^{i \phi S_{\Bal,3}} = e
\labE{gBal2}
}
where the last equality in \refE{gBal1} follows from the definition \Eref{GM} of $S_{\Bal,3}$, and the second line in \refE{gBal2} is derived from $\mr{e}^{ - i \pi S_{\Bal,2}} S_{\Bal,3} \mr{e}^{i \pi S_{\Bal,2}} = - S_{\Bal,3}$. 
It is worthwhile to mention that $\pi_1(G)$ is Abelian, 
which follows from the fact that $L_G$ is Abelian and 
the fact that a quotient group of an Abelian group is Abelian \cite{Robinson96}. 

The second homotopy group of a compact Lie group is known to vanish identically \cite{Cartan36, Hall15, Brocker85}: 
\peqn{
\pi_2(G) \simeq 0
\labE{homG2}
}

We now discuss the third \hg. 
It is known that the Lie algebra $\Gg$ of a compact Lie group $G$ can be decomposed into the direct sum 
of one-dimensional Lie algebras $\uni$ and a set of compact simple Lie algebras $\tbr{\Gg_i}_{i=1}^a$ \cite{Hall15}: 
\ceqn{
\Gg = \uni^{a\pri} \oplus \bigoplus_{i=1}^a \Gg_i
\labE{homm3.01}
}
where $a$ and $a\pri$ are the integers which are uniquely determined from $\Gg$, 
and $\uni$ is the Lie algebra of $\mr{U}(1)$, the unitary group of degree one. 
Let $\Bal_i$, $\Bal_i^c$, and $\bm{S}_{\Bal_i}$ be 
one of the root vectors in $\Gg_i$ with the largest length, 
the corresponding co-root, 
and the corresponding generalized $\su(2)$-spin vector defined in \refE{GM}, respectively. 
We define $g_{\Bal_i}^{(3)}:S^3 \to G$ for $\bm{S}_{\Bal_i}$ by 
\ceqn{
g_{\Bal_i}^{(3)}(\psi, \theta, \phi) &:=& \exp\rbr{2 i \psi \bm{S}_{\Bal_i} \cdot \hat{\bm{r}}(\theta, \phi)}
\labE{hom3.02}
}
where $\hat{\bm{r}}(\theta, \phi)$ is a unit vector on $S^2$ defined by $\hat{\bm{r}}(\theta, \phi) := (\sin\theta\cos\phi,\sin\theta\sin\phi,\cos\theta)$, 
with $\psi, \theta$, and $\phi$ being the polar coordinates of the three-dimensional sphere $S^3$: 
\peqn{
S^3 =&&\{
(\sin\psi\sin\theta\cos\phi, \sin\psi\sin\theta\sin\phi, \sin\psi\cos\theta, \cos\psi)
\nnco
&& \ \ | \psi \in [0,\pi], \theta \in [0,\pi], \phi \in [0,2\pi]
\}
\labE{3dpolarcoordinate}
}
Since $\bm{S}_{\Bal_i} \cdot \hat{\bm{r}}(\theta, \phi)$ is the projection of the generalized $\su(2)$-spin vector in the direction of $\hat{\bm{r}}(\theta, \phi)$, 
$g_{\Bal_i}^{(3)}(\psi, \theta, \phi)$ describes the rotation of $\bm{S}_{\Bal_i}$ 
about $\hat{\bm{r}}(\theta, \phi)$ through angle $2 \psi$ [see \refF{g3operation}]. 
For example, 
$g_{\Bal_i}^{(3)}(\psi, \theta, \phi)$ for $G = \mr{SO}(3)$ 
describes rotation in three dimensions about a vector $\hat{\bm{r}}(\theta, \phi)$ through angle $2 \psi$. 
Let $\rbr{f}_M$ be the homotopy class of $f$ on a manifold $M$, 
which is the set of those maps to $M$ that can continuously transform into $f$, 
where $f$ is referred to as a representative element. 
Then, the following lemma holds. 
\begin{lemm}
\labTH{lemm1}\textit{
The third homotopy group of $G$ is generated by the set $\tbr{\rbr{g_{\Bal_i}^{(3)}}_G }_{i=1}^a$: 
\ceqn{
\pi_3(G) \simeq \tbr{\left. \sum_{i=1}^a m_i \rbr{g_{\Bal_i}^{(3)}}_G \right| m_i \in \mbZ} \simeq \mbZ^a
\labE{homG3}
}
where we denote the product on $\pi_3(G)$ as the sum since $\pi_3(G)$ is Abelian. 
}
\end{lemm}
The proof of \refLM{lemm1} is given in \refA{proof1}. 
We note that the homotopy class $\rbr{g_{\Bal_i}^{(3)}}_G$ does not depend on the choice of the root vector $\Bal_i$ 
since $g_{\Bal_i}^{(3)}$ and $g_{\Bal_i\pri}^{(3)}$ can continuously transform into each other 
if the corresponding root vectors $\Bal_i$ and $\Bal_i\pri$ in $\Gg_i$ both have the largest length \cite{Onishchik94}.

\begin{figure}
\centering
\includegraphics[width=6cm, bb=0 0 360 300]{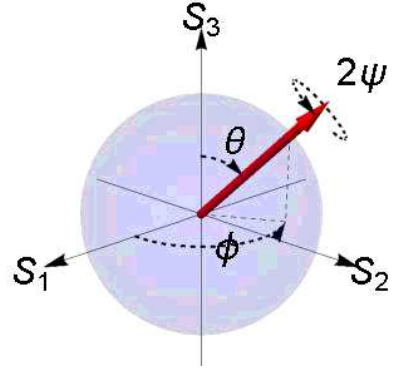}
\caption{\labF{g3operation} 
(Color online) 
Schematic illustration of 
the symmetry transformation $g_{\Bal}^{(3)}(\psi, \theta, \phi)$ 
defined in \refE{hom3.02}. 
The red arrow indicates the generalized $\su(2)$-spin vector parallel to 
the unit vector $\hat{\bm{r}}(\theta, \phi) := (\sin\theta\cos\phi,\sin\theta\sin\phi,\cos\theta)$ 
and $g_{\Bal}^{(3)}(\psi, \theta, \phi)$ describes 
the spin rotation about $\hat{\bm{r}}(\theta, \phi)$ through angle $2 \psi$. 
}
\end{figure}

\subsection{\labS{3.B} Homotopy groups of the order parameter manifold $G/H$}

\subsubsection{\labS{3.B.1} Two types of textures on $G/H$}
Let $D^m$ be the surface and the inner region of an $m$-dimensional sphere $S^m$ with radius $\pi$: 
\ceqn{
D^m := \tbr{\bm{x} \in \mbR^m | \norm{x} \le \pi }
\labE{Dm}
}
and consider a topological excitation without a defect characterized by $\pim$, 
such as two-dimensional $(m=2)$ and three-dimensional $(m=3)$ skyrmions. 
Assuming that it is localized in $D^m$, 
we may regard its texture $O(\bm{x})$ as a map from $D^m$ to the $G/H$ 
subject to the boundary condition 
\ceqn{
O(\bm{x}) = O_0 
\FOR \norm{x} = \pi
\labE{boundarym2}
}
where $O_0$ is a fixed value of the order parameter called the reference order parameter 
and $\norm{x}$ denotes the modulus of $\bm{x}$. 
We note that 
a map $O: D^m \to G/H$ subject to the boundary condition \Eref{boundarym2} can represent 
a texture of a topological excitation with a defect 
through the replacement of $D^m$ by an $m$-dimensional sphere $S^m$ enclosing the defect. 
A crucial point for obtaining $\pim$ is 
to express the texture $O(\bm{x})$ in terms of a symmetry transformation $g(\bm{x})$ 
depending on the space coordinate $\bm{x}$ as follows: 
\ceqn{
O(\bm{x}) = g(\bm{x}) O_0
\labE{actionrep}
}
where $g O_0$ for $g \ (\in G)$ denotes the action of $g$ on $O_0$. 
The expression \Eref{actionrep} relates a texture $O(\bm{x})$ on $G/H$ to a texture $g(\bm{x})$ on $G$ 
and $\pim$ to $\pi_m(G)$ and $\pi_m(H)$. 
Although $O(\bm{x})$ is continuous on $D^m$, 
$g(\bm{x})$ may not be continuous 
because $g(\bm{x})$ and $g(\bm{x})h(\bm{x})$ for any discontinuous function $h$ with $h(\bm{x}) \in H$ give the same texture $O(\bm{x})$ in \refE{actionrep}. 
As we will see below, 
two cases arise depending on whether or not $g(\bm{x})$ is continuous on the entire region of $D^m$. 

Given a subgroup $H$ of $G$, 
we can define the inclusion map $i : H \to G$ by $i(h) := h$ [see \refF{cokerker} (a)]. 
Then, a texture on $H$, i.e., a map $g$ from $D^m$ to $H$, can also be regarded as a texture on $G$, 
and we define a map $i_{\ast m}: \pi_m(H) \to \pi_m(G)$ between the homotopy groups as $i_{\ast m}(\rbr{f}_H) = \rbr{i \circ f}_G$, 
where $\circ$ denotes the composition of two maps. 
We construct textures $G/H$ in two ways from two groups $\CIM$ and $\KIM$ which are defined as follows: 
\ceqn{
\CIM &:=& \COKb{i_{\ast m}: \pi_m(H) \to \pi_m(G)} 
\nnco
&:=& {\pi_m(G) \over \mr{Im}\tbr{i_{\ast m}: \pi_m(H) \to \pi_m(G)}}
\labE{defcim}
\eco
\KIM &:=& \KERb{i_{\ast m-1}: \pi_{m-1}(H) \to \pi_{m-1}(G)} 
\nnco
&:=& \tbr{O \in \pi_{m-1}(H) | i_{\ast m-1}(O) = e}
\labE{defkim}
}
where $\mr{Im} \ F := \tbr{F(g) | g \in G}$ and $\mr{Coker} \ F$ for $F: G \to G\pri$ is defined by $\mr{Coker} \ F := G\pri/ \mr{Im} \ F$ [see \refF{cokerker} (b)]. 
An element of $\mr{Ker} \ i_{\ast m-1}$ represents a nontrivial texture on $H$ 
that is trivial as a texture on $G$. 
While an element of $\mr{Im} \ i_{\ast m}$ represents a nontrivial texture on $G$ 
that can be represented as a texture on $H$, 
that of $\CIM$ represents a nontrivial texture on $G$ 
that cannot be represented as a texture on $H$. 
We denote the element of $\CIM$ corresponding to $a \in \pi_m(G)$ by $\rbr{a}$ 
and call $a$ the representative element of $\rbr{a}$. 

\begin{figure}
\centering
\includegraphics[width=8cm, bb=0 0 703 459]{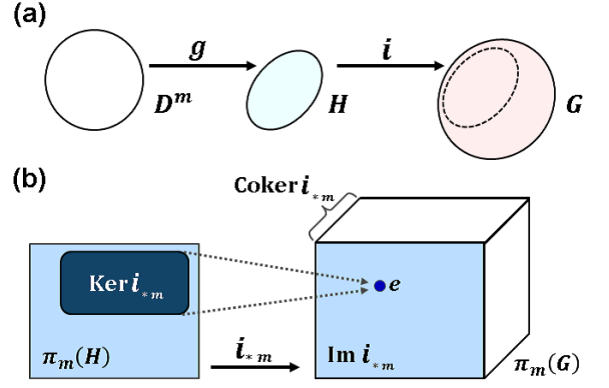}
\caption{\labF{cokerker} 
(Color online) 
(a) 
Schematic illustration of an inclusion map $i: H \to G$. 
Given a map $g$ from $D^m$ to $H$, 
the composition $i \circ g$ gives a map from $D^m$ to $G$, 
where $D^m$ is an $m$-dimensional disk defined in \Eref{Dm}. 
(b) 
Schematic illustration of the cokernel $\CIM$, the kernel $\mr{Ker} \ i_{\ast m}$, and the image $\mr{Im} \ i_{\ast m}$ 
of $i_{\ast,m}$. 
The kernel $\mr{Ker} \ i_{\ast m-1}$ is a subgroup of $\pi_{m-1}(H)$, 
whose elements are mapped to the identity element $e$ of $\pi_{m-1}(G)$. 
The image $\mr{Im} \ i_{\ast m}$ is a subgroup of $\pi_m(G)$, 
whose elements are obtained through $i_{\ast m}$. 
The cokernel $\CIM$ is the quotient group $\pi_m(G)/\mr{Im} \ i_{\ast m}$, 
representing the elements in $\pi_m(G)$ 
that cannot be obtained through $i_{\ast m}$. 
}
\end{figure}

Let us construct the texture $O^{\rbr{a}}$ on $G/H$ from $\rbr{a} \in \CIM$, 
Since $a$ is a texture of $G$, 
we can define the texture $O^{\rbr{a}}$ through the action of $a$ on $O_0$: 
\peqn{
O^{\rbr{a}}(\bm{x}) := a(\bm{x}) O_0 
\FOR \bm{x} \in D^m
\labE{tex.cok}
}
Equation \Eref{tex.cok} implies that 
a nontrivial texture on $G/H$ can be obtained from a nontrivial texture on $G$ [see \refF{construction} (a)]. 
From the boundary condition for $a$, i.e., 
\ceqn{
a(\bm{x}) = e \FOR \norm{x} = \pi
\labE{boundarym22}
}
we see that $O^{\rbr{a}}(\bm{x})$ satisfies the boundary condition \Eref{boundarym2}. 
It is worth mentioning two things. 
First, $\CIM$ is Abelian for $m\ge 1$ 
because the numerator on the right-hand side of \refE{defcim}, i.e., $\pi_m(G)$, is Abelian. 
This follows from the fact that $\pi_1(G)$ is Abelian and 
from the commutativity of higher-dimensional homotopy groups \cite{Nakahara03}. 
Second, we must consider the quotient space $\CIM$ instead of $\pi_m(G)$, 
which is the numerator on the right-hand side of \refE{defcim}, 
because the denominator $\mr{Im} \ i_{\ast m}$ gives a uniform texture through \refE{tex.cok}. 
Indeed, for $a_h := i_{\ast m}(a_H) \in \mr{Im} \ i_{\ast m}$, we have 
\ceqn{
O^{[a_h]}(\bm{x}) &:=& a_h(\bm{x}) O_0 = \rbr{i(a_H)}(\bm{x})O_0 
\nnco
 &=& a_H(\bm{x})O_0 = O_0
\FOR \forall \bm{x} \in D^m
}
where we use the invariance of $O_0$ under the transformation in $H$ in obtaining the last equality. 
The simplest example of the construction \Eref{tex.cok} is an integer-quantum vortex in a scalar BEC. 
Let $\Psi$ be the mean-field wave function of the condensate. 
Then, the texture $\Psi(\phi)$ around a vortex with a unit winding number is given by 
\ceqn{
\Psi(\phi) = \exp(i \phi) \Psi_0
\labE{integerquantumvortex}
}
where $\phi$ and $\Psi_0$ are the azimuth angle around the vortex 
and the value of the mean-field wave function at $\phi = 0$. 
Thus, 
the nontrivial texture $\Psi(\phi)$ of the vortex is expressed in terms of 
the nontrivial winding $\exp(i \phi)$ on the symmetry group $G = U(1)$. 
Another example of the construction \Eref{tex.cok} is a three-dimensional skyrmion in a ferromagnet called a knot soliton \cite{Dzyloshinskii79}, 
whose texture $\bm{M}(r, \theta, \phi)$ of the spin is written as 
\ceqn{
\bm{M}(r, \theta, \phi) := \exp\rbr{2 i \psi(r) \bm{S} \cdot \hat{\bm{r}}(\theta, m \phi)} \bm{M}_0
\labE{knotFerro}
}
where $(r, \theta, \phi)$ is the three-dimensional polar coordinates 
and $m \in \mbZ$ denotes the topological charge of the knot soliton. 
Here, $\bm{M}_0 := (0,0,1)$ and $\psi(r)$ is a function that satisfies $\psi(0) = 0$ and $\psi(\infty) = \pi$. 
One can see from \refE{hom3.02} that the nontrivial texture \Eref{knotFerro} is expressed in terms of 
a nontrivial winding $\exp\rbr{2 i \psi(r) \bm{S} \cdot \hat{\bm{r}}(\theta, \phi)}$ on a symmetry group $\mr{SO}(3)$ of spin rotation.

\begin{figure}
\centering
\includegraphics[width=8cm, bb=0 0 723 507]{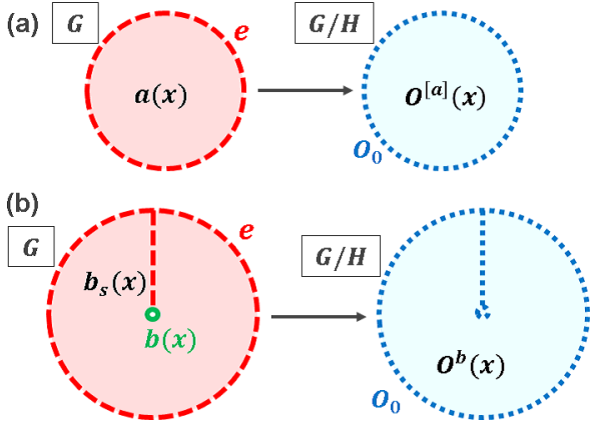}
\caption{\labF{construction} 
(Color online) 
(a) 
Construction of the texture $O^{\rbr{a}}$ from the map $a$ defined in \refE{tex.cok} for $m=2$, 
where $a$ is a map from a two-dimensional disk $D^2$ to $G$ that maps the boundary (red dashed line) of $D^2$ to the identity element $e$. 
The texture $O^{\rbr{a}}$ is a map from $D^2$ to $G/H$  
that maps the boundary (blue dotted line) of $D^2$ to the reference order parameter $O_0$. 
(b) 
Construction of the texture $O^b$ from the map $b$ defined in \refE{tex.ker} for $m=2$. 
Here, $b$ is a map from a circle $S^1$ to $H$ and 
$b_s$ is a continuous deformation from $b$ to the uniform texture $e$ subject to the condition \Eref{BCconti}, 
where the points on the red dashed line are mapped to $e$. 
The texture $O^b$ is a map from $D^2$ to $G/H$ 
which maps the points on the blue dotted line to $O_0$. 
}
\end{figure}

To construct the texture $O^b$ on $G/H$ from $b \in \KIM$, 
we regard $b$ as a map from $S^{m-1}$. 
Since $b$ is a trivial texture on $G$ from its definition \Eref{defkim}, 
there exists a continuous deformation $b_s$ from $b_{s=0} = b$ to the uniform texture $b_{s=\pi} = e$ 
subject to the boundary condition 
\ceqn{
b_s(\hat{\bm{x}}_0) = O_0 
\FOR \forall s \in [0,\pi]
\labE{BCconti}
}
where $\hat{\bm{x}}_0$ is a point on $S^{m-1}$ and $s$ is the parameter of the deformation. 
Hence, we define $O^b$ as 
\ceqn{
O^b(\bm{x}) := b_{s = \norm{x}}\br{ \hat{\bm{x}} } O_0
\FOR \bm{x} \in D^m
\labE{tex.ker}
}
where $\hat{\bm{x}}$ is the unit vector parallel to $\bm{x}$ [see \refF{construction} (b)]. 
We note that 
$b_{s = \norm{x}}\br{\hat{\bm{x}}}$ in the construction \Eref{tex.ker} is not continuous at the origin $\bm{x} = \bm{0}$. 
From the comparison of \refE{actionrep} with \refE{tex.cok} (\refE{tex.ker}), 
a texture on $G/H$ is expressed by a texture on $G$ that is (not) continuous on $D^m$, 
and is described by an element of $\CIM$ ($\KIM$). 
Examples of the construction \Eref{tex.ker} include a half vortex in a uniaxial nematic liquid crystal and a monopole in a ferromagnet. 
The order parameter of a uniaxial nematic liquid crystal is the orientation $\bm{d}$ of molecules. 
The texture $\bm{d}(\phi)$ around a half vortex is given by 
\ceqn{
\bm{d}(\phi) := \exp(\phi L_2/2) \bm{d}_0
\labE{halfvortex}
}
where $\bm{d}_0 := (0,0,1)$, and $\phi$ and $L_2$ are the azimuth angle around the vortex 
and a generator of rotation about the $y$-axis, respectively. 
The nontrivial texture $\bm{d}(\phi)$ is expressed not by a loop on $G = \mr{SO}(3)$ 
but by a path from $e$ to $\exp(\pi L_2)$ on $\mr{SO}(3)$. 
Due to the discrete $\pi$-rotational symmetry $\bm{d} \to - \bm{d}$, 
the start point $\bm{d}_0$ and the end point $\exp(\pi L_2)\bm{d}_0 = -\bm{d}_0$ should be identified, 
where the texture \Eref{halfvortex} is continuous at $\phi =0$ ($\phi = 2\pi$). 
The texture $\bm{M}(\theta,\phi)$ of a monopole in a ferromagnet 
is described by a hedgehog configuration of the spin, i.e., $\bm{M}(\theta,\phi) = \hat{\bm{r}}(\theta, \phi)$, 
which can be rewritten in terms of the $\su(2)$-spin vector $\bm{S}$ as follows: 
\ceqn{
\bm{M}(\theta,\phi) := \exp(i \phi S_3)\exp(i \theta S_2)\bm{M}_0
\labE{simplemon}
}
where $\theta$ and $\phi$ denote the polar angle and the azimuth angle around the monopole, respectively, 
$\bm{M}_0 := (0,0,1)$, and $S_2$ ($S_3$) is a generator of the rotation about the $y$-axis ($z$-axis). 
As shown in \refF{monopole}, 
the hedgehog texture is obtained by 
the successive applications of spin rotation $\exp(i \theta S_2)$ about the $y$-axis 
followed by spin rotation $\exp(i \phi S_3)$ about the $z$-axis. 
Under continuous deformation $\bm{M}_u(\theta,\phi) =\exp(-i u\theta S_2) \exp(i \phi S_3)\exp(i \theta S_2) \bm{M}_0$, 
with $u \ (\in [0,1])$ being the parameter of the deformation, 
$\bm{M}_{u=0}(\theta,\phi) = \bm{M}(\theta,\phi)$ transforms into 
\ceqn{
&& \bm{M}_{u=1}(\theta,\phi) 
\nnco
&=& \exp(- i \theta S_2) \exp(i \phi S_3)\exp(i \theta S_2) \bm{M}_0
\nnco
&=& \exp(- i \theta S_2) \exp(i \phi S_3)\exp(i \theta S_2) \exp(i \phi S_3) \bm{M}_0
\nnco
&=& g^{(2)}(\theta,\phi) \bm{M}_0
}
where in the second equality we use the invariance of $\bm{M}_0$ under the rotation about the $z$-axis, 
and $g^{(2)}(\theta,\phi) := \exp(- i \theta S_2) \exp(i \phi S_3)\exp(i \theta S_2) \exp(i \phi S_3)$ 
is the map \Eref{deform} with $\bm{S}_{\Bal}$ replaced by $\bm{S}$. 
The expression $g^{(2)}(\theta,\phi) \bm{M}_0$ 
gives the texture of a monopole in the form of \refE{actionrep}. 
Indeed, since we have 
\ceqn{
g^{(2)}(\theta= 0,\phi) &=& \exp(2 i \phi S_3) 
\eco
g^{(2)}(\theta= \pi,\phi) &=& \mr{e}^{- i \pi S_2} \mr{e}^{i \phi S_3} \mr{e}^{i \pi S_2} \mr{e}^{i \phi S_3} 
\nnco
&=& \mr{e}^{-i \phi S_3} \mr{e}^{i \phi S_3} = e
\eco
g^{(2)}(\theta, \phi= 0) &=& e
}
where $g^{(2)}(\theta, \phi)$ is a continuous deformation from the loop $g^{(2)}(\theta= 0,\phi) = \exp(2 i \phi S_3)$ on $H = \mr{SO}(2)$ 
to the trivial loop $g^{(2)}(\theta = \pi, \phi) = e$ subject to the boundary condition \Eref{BCconti}. 

\begin{figure}
\centering
\includegraphics[width=8cm, bb=0 0 552 522]{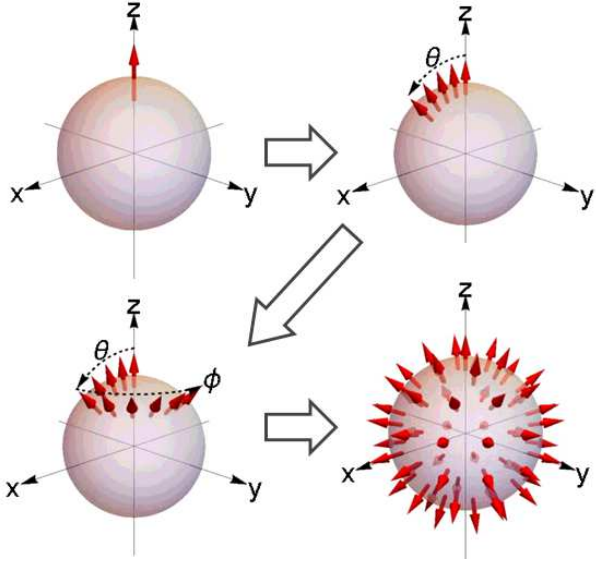}
\caption{\labF{monopole} 
(Color online) 
Schematic illustration of a monopole in a ferromagnet with \tcha \ $m = +1$. 
Each arrow indicates the direction of the spin. 
The texture of the monopole is described by a hedgehog configuration of the spin, 
which is obtained from successive applications of spin rotation $\exp(i \theta S_2)$ about the $y$-axis through angle $\theta$ 
followed by spin rotation $\exp(i \phi S_3)$ about the $z$-axis through angle $\phi$, 
where $\theta$ and $\phi$ are the polar angle and the azimuth angle, respectively. 
}
\end{figure}

Finally, we define the map 
\neqn{
f: \KIM \times \KIM \to \CIM
} 
which is used in \refTH{theo1} below. 
As shown in \refS{4}, $f$ vanishes for $m=2$ and $3$, 
so long as we focus on the first, second, and third homotopy groups. 
We therefore first define $f$ for $m=1$ and then define it for an arbitrary dimension $m$. 
For $m=1$, 
we have the following isomorphism: 
\ceqn{
\KER i_{0\ast} \simeq \dis
\labE{KIM0}
}
where $G_0$ is a Lie group constituted from the connected component in $G$.
This follows from the fact that $[h] \in \KER i_{0\ast}$ implies that 
$h$, a representative element of $[h]$, is connected to $e$ by a path on $G$, 
which implies $h \in G_0$, and 
from the fact that $[h] \in \dis$ describes a common element of $\pi_0(H)$ and $\pi_0(G)$. 
Let $[\sig]$ and $[\tau]$ be elements of $\dis$ 
with representative elements $\sig \ (\in H \cap G_0)$ and $\tau \ (\in H \cap G_0)$, respectively. 
Since they are connected with the identity element $e$, 
there exist paths $\gamma_\sig, \gamma_\tau$, and $\gamma_{\sig\tau}$ from $\sig, \tau$, and $\sig\tau$, respectively, to $e$. 
Then, the composition $\gamma_\sig  \gamma_\tau  \br{\gamma_{\sig\tau}}^{-1}$ is a loop from $e$ to itself [see \refF{factorset}]. 
We define $f$ as 
\peqn{
f([\sig],[\tau]) := \gamma_\sig  \gamma_\tau  \br{\gamma_{\sig\tau}}^{-1} O_0
\labE{factorsets1}
}
Let $b$ and $b\pri$ be elements of $\KIM$. 
Then, there exist continuous deformations $b_s$, $b\pri_s$, and $(bb\pri)_s$ from $b$, $b\pri$, and $bb\pri$, respectively, to a trivial map subject to the boundary condition \Eref{BCconti}. 
We fix the parameter $s$, 
consider $b_s$ and $b\pri_s$ to be elements of $\pi_{m-1}(G)$, 
and denote their composition as $b_s \circ b_s\pri$. 
Then, we define $f(b,b\pri)$ by 
\neqn{
&& [f(b,b\pri)](\bm{x}) := \rbr{\widetilde{f}(b,b\pri)}(\bm{x}) O_0 
\eco
&& \rbr{\widetilde{f}(b,b\pri)}(\bm{x}) 
\nnco
&& := 
\begin{cases}
(bb\pri)_{s = \pi - 2 \norm{x}} \br{ \hat{\bm{x}} } 
&\FOR 0 \le \norm{x} \le \frac{\pi}{2}
;\\
(b_{s= 2 \norm{x} - \pi} \circ b_{s= 2 \norm{x} - \pi}\pri) ( \hat{\bm{x}} ) 
&\FOR \frac{\pi}{2} \le \norm{x} \le \pi
. 
\end{cases}
\nnco
\labE{factorsetsm}
}
Since $\widetilde{f}$ satisfies the boundary condition \Eref{boundarym22}, 
which gives $\widetilde{f} \in \pi_m(G)$, 
$f$ is indeed a map to $\CIM$. 
The map $f$ does not depend on the choices of the representative elements of $b$ and $b\pri$. 
Let $B$ and $\bar{B}$ be two representative elements of $b$, 
and $B_s \ (\bar{B}_s)$ be the continuous deformations from $B \ (\bar{B})$ to the trivial homotopy class. 
Since $B$ and $\bar{B}$ transform into each other through continuous deformation on $H$, 
so do $B_s$ and $\bar{B}_s$. 
Therefore, the map $f$ in \refE{factorsetsm} defined from $B$ and $B_s$ 
and that defined from $\bar{B}$ and $\bar{B}_s$ transform into each other continuously. 

\begin{figure}
\centering
\includegraphics[width=6cm, bb=0 0 566 425]{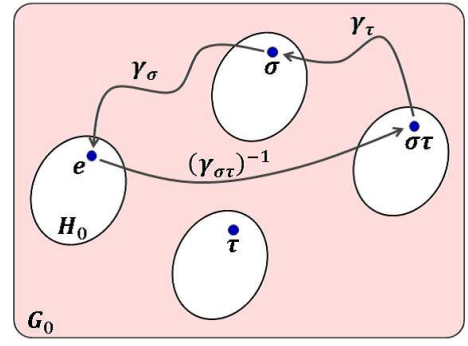}
\caption{\labF{factorset} 
(Color online) 
Schematic illustration of the map $f$ defined in \refE{factorsets1}. 
The entire region shows the connected component $G_0$ of $G$ 
and the white regions represent the connected components of $H$, 
where $H_0$ shows the connected component including the identity element $e$, 
and $\sig$, $\tau$, and $\sig\tau$ denote the elements of $G_0$ 
in different connected components in $H$. 
The map $f$ is obtained from the composition $\gamma_\sig  \gamma_\tau  \br{\gamma_{\sig\tau}}^{-1}$ of the three paths $\gamma_\sig$, $\gamma_\tau$, and $\gamma_{\sig\tau}$. 
}
\end{figure}

\subsubsection{\labS{3.B.2} A decomposition formula for $\pim$}
Let us define the product $\times_f$ on the product set $\CIM \times \KIM$ by 
\ceqn{
([a], b) \prf ([a\pri],b\pri) := 
([a]+[a\pri]+f(b,b\pri), bb\pri)
\labE{homm3}
}
where $f: \KIM \times \KIM \to \CIM$ is the map defined in \refE{factorsetsm} 
and we denote the product in $\CIM$ as the sum since $\CIM$ is Abelian. 
Then the following theorem holds. 

\begin{theo}
\labTH{theo1}\textit{
Under the product defined in \refE{homm3}, $\CIM \times \KIM$ becomes a group. 
This group denoted by $\CIM \prf \KIM$ is isomorphic to the $m$th homotopy group of $G/H$: 
\peqn{
\pim \simeq \CIM \prf \KIM
\labE{homm}
}
Any topological charge $([a],b)$ in $\pim$ can be uniquely decomposed into the product of an element of $\CIM$ and that of $\KIM$: 
\peqn{
&&([a], b) = ([a],e) \prf (e,b)
\labE{homm4} 
}
Furthermore, the texture $O^{\rbr{a}}$ $(O^b)$ of a \te \ with \tcha \ $([a],e)$ $((e,b))$ is given by \refErm{tex.cok} $($\refErm{tex.ker}$)$. 
}
\end{theo}
The proof of \refTH{theo1} is given in \refA{proof2}. 
Equation~\Eref{homm} implies that 
there are two distinct types of \te s expressed by either $\CIM$ or $\KIM$. 
One can see from \refE{homm4} that 
any \te \ can be written as a composition of these two types. 
The presence of $f$ in \refE{homm3} implies that these two types of \te s are, in general, not independent: 
the composition of two \te s described by $\KIM$ can produce a \te \ described by $\CIM$, 
because we have 
\ceqn{
(e, b) \times_f (e, b\pri) &=& (f(b,b\pri), bb\pri) 
\nnco
&=& (f(b,b\pri), e) \times_f (e, bb\pri)
}
and $f(b,b\pri) \ne e$ in general. 
The group $\CIM \prf \KIM$ is referred to as the group extension of $\CIM$ by $\KIM$ with the factor set $f$ \cite{Robinson96} (see \refA{B.1} or Ref.~\cite{Robinson96} for detail).

\section{\labS{4} Formulas for homotopy groups for low-dimensional topological excitations}

\subsection{\labS{4.A} First homotopy group: vortices}
Since $\COKo$ is a quotient group of $\pog$ from \refE{defcim} and $\pog$ is a quotient group of $L_G$ from \refE{integrallattice}, 
we write an element of $\COKo$ as $[H_{\bm{t}}]$ where $H_{\bm{t}} \in L_G$.  Let us define the product $\times_f$ on $\COKo \times \dis$ by 
\ceqn{
&&(\rbr{H_{\bm{t}}},[\sig]) \prf (\rbr{H_{\bm{s}}}, [\tau]) 
\nnco
 &:=& (\rbr{H_{\bm{t}}} + \rbr{H_{\bm{s}}} + f([\sig],[\tau]), [\sig][\tau])
\labE{hom1.2}
}
where $f$ is the map defined in \refE{factorsets1}. 
Then the following corollary holds. 
\begin{coro}
\labTH{cor1}\textit{
Under the product defined in \refE{hom1.2}, $\COKo \prf \dis$ is isomorphic to the first homotopy group of $G/H$: 
\peqn{
\po \simeq \COKo \prf \dis
\labE{hom1}
}
Any topological charge $(\rbr{H_{\bm{t}}},[\sig])$ in $\po$ can be uniquely decomposed into the product of 
an element of $\COKo$ and that of $\dis$:
\peqn{
(\rbr{H_{\bm{t}}},[\sig]) = (\rbr{H_{\bm{t}}},e) \prf (e,[\sig])
\labE{hom1.3}
}
Let $\phi$ and $O_0$ be the azimuth angle around a vortex and the reference order parameter, respectively. 
The texture $O^{(\rbr{H_{\bm{t}}},e)}(\phi)$ of a vortex with \tcha \ $(\rbr{H_{\bm{t}}},e)$ is given by 
\ceqn{
O^{(\rbr{H_{\bm{t}}},e)}(\phi) = \exp(i \phi H_{\bm{t}}) O_0
\labE{hom1.1.1}
}
while the texture $O^{(e,\rbr{\sig})}(\phi)$ of a vortex with \tcha \ $(e, \rbr{\sig})$ is given by 
\ceqn{
O^{(e,\rbr{\sig})}(\phi) = \gamma_{\sig}(\phi) O_0
\labE{hom1.1.2}
}
where $\gamma_{\sig}(\phi)$ is a path from $\sig \ (\in H \cap G_0)$ to the identity element $e$. 
}
\end{coro}
\textit{Proof}\\
Due to the isomorphism $\KER i_{0\ast} \simeq \dis$ in \refE{KIM0}, 
we have Eqs.~\Eref{hom1} and \Eref{hom1.3} from Eqs.~\Eref{homm} and \Eref{homm4}, respectively. 
Then, it is sufficient to show Eqs.~\Eref{hom1.1.1} and \Eref{hom1.1.2} to prove \refCO{cor1}. 
From \refE{tex.cok}, 
the texture of a vortex with \tcha \ $([H_{\bm{t}}], e)$ is given by $O^{([H_{\bm{t}}], e)}(\phi) = a(\phi) O_0$, 
where $a(\phi)$ is a loop on $G$. 
From Eqs.~\Eref{homG1} and \Eref{homG1.1}, 
we have $a(\phi) = \exp(i \phi H_{\bm{t}})$ and hence \refE{hom1.1.1}. 
For a vortex with \tcha \ $(e, \rbr{\sig})$, 
\refE{tex.ker} can be expressed as $O^{(e,[\sig])}(\phi) = b_\phi O_0$, 
where $b$ is a path from $b_{s = 0} = \sig$ to $b_{s = \pi} = e$. 
Defining a path $\gamma_\sig$ by $\gamma_\sig(\phi) := b_{s = \phi/2}$, 
we obtain \refE{hom1.1.2}, 
which completes the proof of \refCO{cor1}. 

Equation~\Eref{hom1} implies that 
there are two types of vortices expressed by either $\COKo$ or $\dis$, 
and it follows from \refE{hom1.3} that 
any vortex can be written as their composition. 
Examples of the former include an integer-quantum vortex \Eref{integerquantumvortex}, 
which is obtained from \refE{hom1.1.2} through substitution of $\exp( i \phi H_{\bm{t}} )$ and $O^{(\rbr{H_{\bm{t}}},e)}(\phi)$ with 
$\exp(i \phi)$ and $\Psi(\phi)$, respectively.
The latter term $\dis$ describes 
a vortex associated with a discrete symmetry of the state such as a half vortex \Eref{halfvortex} in a uniaxial nematic liquid crystal, 
where the discrete symmetry is the $\pi$-rotational symmetry of the orientation $\bm{d}$. 
One can reproduce from \refE{hom1} the formula $\po \simeq \pi_0(\widetilde{H})$ based on the lift method \cite{Mermin79}, 
where $G$ and $H$ are lifted to a simply connected group $\widetilde{G}$ and the corresponding subgroup $\widetilde{H}$, respectively. 
Since $\pi_0(\widetilde{G}) \simeq 0$ and $\pi_1(\widetilde{G}) \simeq 0$, 
we have $\COKo \simeq 0$ and $\pi_0(\widetilde{H} \cap \widetilde{G}_0) \simeq \pi_0(\widetilde{H})$. 
We thus obtain \refE{hom1}. 
In contrast to the lift method, we find the distinction between vortices represented by $\COKo$ and $\dis$. 
In \refS{5.A}, 
this distinction is shown to be crucial since only the latter can be a cause of nontrivial \tinf.

\subsection{\labS{4.B} Second homotopy group: monopoles and skyrmions}
We generalize the texture \Eref{simplemon} of a monopole in a ferromagnet 
through the replacement of $\bm{S}$ by a generalized $\su(2)$-spin vector. 
Provided that $S_{\Bal,3}$ is an unbroken generator, 
we define the mapping $O^{\Bal^c}: S^2 \to G/H$ for a co-root $\Bal^c$ by 
\peqn{
O^{\Bal^c}(\theta,\phi) &:=& g_{\Bal}^{(2)}(\theta,\phi) O_0
\labE{tex.2dSky}
\eco
g_{\Bal}^{(2)}(\theta,\phi) &:=& \exp(i \phi S_{\Bal,3})\exp(i \theta S_{\Bal,2})
\labE{tex.2dSky.1}
}
Comparing Eqs.~\Eref{tex.2dSky} and \Eref{tex.2dSky.1} with \refE{simplemon}, 
we find that \refE{tex.2dSky} describes the hedgehog configuration of the generalized $\su(2)$-spin vector $\bm{S}_{\Bal}$. 
More precisely, 
$O^{\Bal^c}(\theta,\phi)$ is invariant under 
spin rotations generated by the generalized $\su(2)$-spin vector parallel to $\hat{\bm{r}}(\theta, \phi)$: 
\neqn{
\exp\rbr{i \psi \bm{S}_{\Bal} \cdot \hat{\bm{r}}(\theta, \phi) }O^{\Bal^c}(\theta,\phi) = O^{\Bal^c}(\theta,\phi) \FOR \forall \psi \in \mbR
. \nnco
\labE{tex.2dSky.2}
}
This follows from the assumption that 
$O_0$ is invariant under unitary transformations generated by $S_{\Bal,3}$ 
and from the decomposition $\exp\rbr{i \psi \bm{S}_{\Bal} \cdot \hat{\bm{r}}(\theta, \phi) } = 
g_{\Bal}^{(2)}(\theta,\phi) \mr{e}^{i \psi S_{\Bal,3}} \rbr{ g_{\Bal}^{(2)}(\theta,\phi) }^\dagger$, 
which is derived directly from the commutation relations \Eref{comm}. 

The following \refCO{cor2} shows that 
the \tcha \ and the texture of a general monopole are described 
by co-roots and the hedgehog configuration of the generalized $\su(2)$-spin vector, 
respectively. 
Reflecting the fact that a general compact Lie algebra includes more than one $\su(2)$-Lie algebra in contrast to $\su(2)$, 
the \tcha \ should be described by a set of co-roots. 
The connection with the co-roots and the generalized $\su(2)$-spin vectors are 
pointed out in Refs.~\cite{Weinberg80, Weinberg07}, 
where non-Abelian gauge theories are considered and $H$ is assumed to include a maximal Abelian subgroup of $G$ \cite{Brocker85, Hall15}. 
We here generalize their results to arbitrary systems with arbitrary patterns of symmetry breaking.  

\begin{coro}
\labTH{cor2}\textit{
Let $L_H$ and $L_H^c$ $(L_G^c)$ be the integral lattice of $H$ and the co-root lattice of $H$ $(G)$, respectively. 
Then, $L_H^c$ is an Abelian subgroup of $L_H \cap L_G^c$, 
and the quotient space of $L_H \cap L_G^c$ by $L_H^c$ 
is isomorphic to $\pt$: 
\peqn{
\pt \simeq (L_H \cap L_G^c)/L_H^c
\labE{hom2}
}
Therefore, the \tcha \ $n$ of a \mon \ can be expressed in terms of the co-roots corresponding to the simple roots as 
\ceqn{
n = \sum_{j=1}^r m_j \Bal_j^c
\labE{hom22}
}
where $\tbr{m_j}_{j=1}^r$ is the set of integers, 
and its texture $O(\theta,\phi)$ is given by 
\neqn{
O(\theta,\phi) = g_{\Bal_1}^{(2)}(\theta,m_1\phi)g_{\Bal_2}^{(2)}(\theta,m_2\phi)\cdots g_{\Bal_r}^{(2)}(\theta,m_r\phi) O_0
. \nnco
\labE{hom23}
}
}
\end{coro}

The proof of \refCO{cor2} is given in \refA{proofcor2}. 
One can reproduce from \refE{hom2} the formula $\pt \simeq \pi_1(\widetilde{H})$ based on the lift method \cite{Mermin79}. 
Indeed, we have $\pi_1(\widetilde{G}) \simeq 0$ and hence $L_{\widetilde{G}} \simeq L_{\widetilde{G}}^c$. 
Then, we obtain $\pt \simeq \pi_1(\widetilde{H})$. 
However, the texture of each \te \ is described by a deformable loop on $\widetilde{G}$; 
in Ref.~\cite{Mermin79} the existence of the texture is shown but no explicit form is given. 
We here explicitly determine the texture as shown in \refE{hom23}.

\subsection{\labS{4.C} Third homotopy group: three-dimensional skyrmions}
Two prototypical examples of \thsky s are 
a Shankar skyrmion and a knot soliton, 
which are characterized by the homotopy groups $\pi_3(S^3) \simeq \mbZ$ and $\pi_3(S^2) \simeq \mbZ$, respectively. 
Both of their textures are expressed in terms of the $\su(2)$-spin vector $\bm{S}$ as 
\ceqn{
O(\psi, \theta, \phi) = \exp\rbr{2 i \psi \bm{S} \cdot \hat{\bm{r}}(\theta, m\phi)} O_0
\labE{unifiedform}
}
where $(\psi, \theta, \phi)$ is the polar coordinates \Eref{3dpolarcoordinate} on $S^3$ 
and $m \in \mbZ$ denotes the \tcha \ of the \thsky. 
This unified description is based on the isomorphism $\pi_3(S^3) \simeq \pi_3(S^2)$ derived from the Hopf fibration \cite{Nakahara03, Manton04, Brocker85}. 
When all of the generators in $\bm{S}$ are broken, 
\refE{unifiedform} describes a Shankar skyrmion \cite{Khawaja01b, Ueda10}; otherwise it describes a knot soliton \cite{Kawaguchi08, Hall16}. 
We generalize the texture \Eref{unifiedform} 
through the replacement of $\bm{S}$ by $\bm{S}_{\Bal}$, 
and define the mapping $O^{\Bal^c}: S^3 \to G/H$ for co-root $\Bal^c$ by 
\ceqn{
O^{\Bal^c}(\psi, \theta, \phi) &:=& g_{\Bal}^{(3)}(\psi, \theta, \phi) O_0
\labE{tex.3dSky}
\eco
g_{\Bal}^{(3)}(\psi, \theta, \phi) &:=& \exp\rbr{2 i \psi \bm{S}_{\Bal} \cdot \hat{\bm{r}}(\theta, \phi)}
\labE{tex.3dSky.1}
}
where $g_{\Bal}^{(3)}$ is defined in \refE{hom3.02}. 

The following two corollaries show that 
a general \thsky \ may be regarded as the composition of several different types of \thsky s 
whose \tcha s and textures are described by co-roots and the corresponding textures \Eref{tex.3dSky}, respectively. 

\begin{coro}
\labTH{cor3}\textit{
The third \hg \ $\pthgh$ is given as follows:
\peqn{
\pthgh \simeq \COKb{i_3^\ast : \pthh \to \pthg}
\labE{hom3}
}
}
\end{coro}
\textit{Proof}\\
Since any subgroup $H$ of a compact Lie group $G$ is compact, $\pi_2(H)$ vanishes. 
Therefore, we obtain $\KER i_2^\ast \simeq 0$ and hence \refE{hom3} from \refTH{theo1}, 
which completes the proof of \refCO{cor3}. 

We next analyze a \tcha \ and a texture. 
Let $\Bal_i^c$ be a co-root of the Lie algebra $\Gg_i$ defined in \refE{homm3.01}. 
Since the numerator $\pi_3(G)$ of \refE{hom3} is generated by $\tbr{ \rbr{g_{\Bal_i^c}^{(3)} } }_{i=1}^a$ from \refLM{lemm1}, 
the quotient space $\pthgh$ is generated by $\tbr{ \rbr{O^{\Bal_i^c}}_{G/H} }_{i=1}^{\bar{a}}$ for a suitable choice of the subset $\tbr{\Bal_k}_{k=1}^{\bar{a}}$ of $\tbr{\Bal_i}_{i=1}^a$. 
Thus, we obtain the following corollary. 

\begin{coro}
\labTH{cor3.1}
\textit{
The \tcha \ $n$ of a \thsky \ can be written in terms of co-roots as 
\ceqn{
n = \sum_{k=1}^{\bar{a}} m_k [\Bal_k^c]
\labE{hom32}
}
where $\tbr{m_k}_{k=1}^r$ is a set of integers 
and $[\Bal^c]$ represents the topological charge of the texture $O^{\Bal^c}$ defined in \refErm{tex.3dSky}. 
}
\end{coro}

The results of this section are summarized in \refT{classify}. 

\begin{table}[t]
\caption{\labT{classify}
Two types of topological excitations and their examples. 
$\CIM$ and $\KIM$ are the cokernel of $i_{\ast m}$ and the kernel of $i_{\ast m-1}$ defined in Eqs.~\Eref{defcim} and \Eref{defkim}, respectively. 
The entry ``absent" means the absence of examples. 
}
\begin{ruledtabular}
\begin{tabular}{c|c|c}
 & $\CIM$ & $\KIM$ 
\\ \hline
$m=1$ & integer-quantum vortex & half vortex
\\ 
$m=2$ & absent & monopole 
\\ 
$m=3$ & knot soliton & absent 
\\
 & Shanker skyrmion &
\\ 
\end{tabular}
\end{ruledtabular}
\end{table}

\section{\labS{5} General conditions for the presence of topological influence}
\subsection{\labS{5.A} Topological influence on a general topological excitation}
When a \te \ with \tcha \ $n \in \pim$ makes a complete circuit of a vortex with \tcha \ $l \in \po$, 
the resulting \tcha \ $\lam_m^l(n)$ is given by the action of $l$ on $n$, 
where the corresponding texture $O^{\lam_m^l(n)}(\bm{x})$ is defined as follows \cite{Kobayashi12a, Kobayashi14a}: 
\neqn{
O^{\lam_m^l(n)}(\bm{x}) := 
\begin{cases}
O^n\br{ 2 \bm{x}}
&\FOR 0 \le \norm{x}\le{\pi\over 2}
; \\
O^l\br{4 \norm{x} - 2 \pi}
&\FOR {\pi\over 2} \le \norm{x}\le \pi
, 
\end{cases}
\nnco
\labE{changem}
}
where $\bm{x} \in D^m$ and $O^n: D^m \to G/H$ ($O^l: [0,2\pi] \to G/H$) is the texture of a topological excitation (vortex) with \tcha \ $n$ ($l$). 
We can express the \tcha s $n$ and $l$ as $n =([a],b)$ and $l = ([H_{\bm{t}}], [\sig])$ from \refTH{theo1} and \refCO{cor1}, respectively, 
Then, the following theorem holds. 
\begin{theo}
\labTH{theo2}\textit{
The \tcha \ $\lam_m^l(n)$ is given by  
\ceqn{
\lam_m^l(n) = ([a], \sig^{-1} b \sig)
\labE{5.1.1}
}
where the homotopy class $\sig^{-1} b \sig$ is defined as $\rbr{\sig^{-1} b \sig}(\bm{x}) := \sig^{-1} b(\bm{x}) \sig O_0$ for $\bm{x} \in D^{m-1}$. 
}
\end{theo}

\begin{table}[t]
\caption{\labT{influence}
Topological influence in four combinations of \te s and vortices. 
Here $\pim$ and $\dis$ are the $m$th and zeroth homotopy groups of the order parameter manifold $G/H$ and $H \cap G_0$, respectively; 
$i_{\ast m}: \pi_m(H) \to \pi_m(G)$ is a homomorphism induced by the inclusion map $i: H \to G$; 
$\CIM$ and $\KIM$ are the cokernel of $i_{\ast m}$ and the kernel of $i_{\ast m-1}$ 
defined in Eqs.~\Eref{defcim} and \Eref{defkim}, respectively. 
The entry ``may appear" (``absent") means that \tinf \ may exist (does not exist). 
}
\begin{ruledtabular}
\begin{tabular}{cc|c|c}
\multicolumn{2}{c|}{} & \multicolumn{2}{c}{\te \ $\pim$}
\\ 
\multicolumn{2}{c|}{}& $\CIM$ & $\KIM$ 
\\ \hline
vortex & $\COKo$ & absent & absent
\\ \cline{2-4}
$\po$ & $\dis$ & absent & may appear
\\ 
\end{tabular}
\end{ruledtabular}
\end{table}

The proof of \refTH{theo2} is given in \refA{prooftheo2}. 
The result of \refTH{theo2} is summarized in \refT{influence}. 
As shown in \refT{influence}, 
only vortices characterized by discrete symmetries can have nontrivial \tinf. 
To understand this, 
let us consider a situation in which a \te \ with texture $O(\bm{x})$ makes a complete circuit of a vortex with texture $O^{(e,[\sig])}(\phi) = \gamma_\sig(\phi) O_0$, 
where $\gamma_\sig(\phi) \sig^{-1}$ describes a path from $e$ to $\sig^{-1}$. 
When the former goes around the latter by angle $\phi$, 
it undergoes a nontrivial texture produced by the latter, 
changing its texture from $O(\bm{x})$ to $\gamma_\sig(\phi)\sig^{-1}O(\bm{x})$. 
The final texture is given by $\sig^{-1} O(\bm{x})$. 
A crucial observation here is that 
the final texture $\sig^{-1} O(\bm{x})$, in general, does not coincide with the initial one $O(\bm{x})$. 
On the other hand, when the topological excitation goes around a vortex characterized by $\COKo$ by angle $\phi$, 
its texture changes from $O(\bm{x})$ to $\exp(i \phi H_{\bm{t}})O(\bm{x})$. 
Therefore, the initial and final textures coincide because we have $\exp(2 \pi i H_{\bm{t}}) = e$ from \refE{integrallattice}.

\subsection{\labS{5.B} Topological influence on low-dimensional topological excitations}

\subsubsection{\labS{5.B.1} Topological influence on a vortex}
The necessary and sufficient condition for the presence of \tinf \ on a vortex 
is the non-Abelianness of the first \hg \ \cite{Kleman77}. 
It is known that non-Abelian vortices behave differently from Abelian ones 
in the collision dynamics \cite{Poenaru77, Kobayashi09, Borgh16}, quantum turbulance \cite{Kobayashi16}, and the coarsening dynamics \cite{Priezjev02, Mcgraw98, Spergel97} 
due to the tangling between vortices. 
However, the conditions for their appearances are yet to be understood from a unified point of view. 
The following corollary shows that 
their presence is solely determined by discrete symmetries, 
where the non-Abelian property is shown to emerge only between pairs of vortices characterized by $\dis$.

\begin{coro}
\labTH{cor4}\textit{
The first homotopy group $\po$ is Abelian if and only if $\dis$ is Abelian and 
$f$ defined in \refErm{factorsets1} satisfies
\peqn{
f([\sig], [\tau]) = f([\tau], [\sig])
\FOR \forall [\sig], [\tau] \in \dis
\labE{cor4.1}
}
}
\end{coro}
\textit{Proof}\\
Comparing the following two equations 
\neqn{
&&([a], [\sig]) \prf ([b], [\tau]) = ([a]+[b]+f([\sig],[\tau]), [\sig][\tau])
, \nnco
&&([a], [\tau]) \prf ([a], [\sig]) = ([a]+[b]+f([\tau],[\sig]), [\tau][\sig])
, \nnco
\labE{cor4.1.2}
}
we find that $\po$ is Abelian if and only if 
\neqn{
\begin{cases}
[\sig][\tau] =[\tau][\sig] 
; & \\
f([\sig], [\tau]) = f([\tau], [\sig]) 
 &
\end{cases}
\FOR \forall [\sig], [\tau] \in \dis
. \nnco
\labE{nonAbelianvortex}
}
The first equation in \refE{nonAbelianvortex} implies that $\dis$ is Abelian 
and we have \refE{cor4.1} from the second equation of \refE{nonAbelianvortex}, 
which completes the proof of \refCO{cor4}.

\subsubsection{\labS{5.B.2}Topological influence on a monopole, a skyrmion, and a \thsky}
Since one \tcha \ changes into another due to \tinf, 
$\lam_2^l$ is an automorphism on $\pt$ \cite{Mermin79, Kobayashi14a}, 
i.e., a one-to-one map from $\pt$ to itself satisfying the homomorphic relation $\lam_2^l(nn\pri) = \lam_2^l(n)\lam_2^l(n\pri)$. 
Therefore, topological influence is characterized by the action of the automorphism group $\mathcal{G}_2$ on $\pt$ defined by 
\peqn{
\mathcal{G}_2 := \tbr{ \lam_2^l | l \in \po}
\labE{autogroup2}
}
From \refCO{cor2}, $\pt$ is described by a co-root lattice $L_G^c$. 
Let us define the Weyl reflection $w_{\Bal}: L_G^c \to L_G^c$ for $\Bal \in R_+$ by 
\ceqn{
w_{\Bal}(H_{\bm{t}}) := H_{\bm{t}\pri}, \ \bm{t}\pri := \bm{t} - {2 (\Bal, \bm{t}) \over (\Bal, \Bal) } \Bal 
\labE{weylreflection}
}
where $w_{\Bal}$ describes the reflection across the plane perpendicular to $\Bal$. 
It is known that $w_{\Bal}$ is an automorphism of $L_G^c$ \cite{Hall15, Brocker85}. 
The Weyl group $W_G$ of $G$ is defined as the automorphism group of $L_G^c$ generated by the Weyl reflections: 
\ceqn{
W_G := \mr{Gen}\tbr{w_{\Bal} | \Bal \in R_+}
}
where $\mr{Gen} \ S$ for a set $S$ is defined as the group generated by the elements of $S$. 
It is instructive to consider an example of $\Gg = \su(2)$. 
Since $\Gg$ is constituted from only one $\su(2)$-subalgebra, 
its co-root lattice is a one-dimensional lattice $L_G^c = \tbr{ m H_{\Bal^c} | m \in \mbZ}$. 
The Weyl reflection acts on $L_G^c$ as its inversion: $w_{\Bal}\br{m H_{\Bal^c}} = - m H_{\Bal^c}$; thus, $W_G \simeq \mbZ_2$. 
More generally, the Weyl group for $\Gg = \su(N)$ is given by $W_G \simeq S_N$, 
where $S_N$ denotes the permutation group of $N$ elements \cite{Hall15, Brocker85}. 
For $\Gg = \uni$, $W_G \simeq 0$, since $\uni$ does not have a co-root. 
From \refCO{cor2}, 
we write an element of $\pt \simeq (L_H \cap L_G^c)/L_H$ by $H_{\bm{t}} + L_H^c$. 
The following corollary shows that \tinf \ is described by a Weyl reflection \Eref{weylreflection} 
and that possible forms of the automorphism group \Eref{autogroup2} are restricted from the Weyl group $W_G$. 
\begin{coro}
\labTH{cor5}\textit{
For each discrete symmetry $\rbr{\sig} \in \dis$, 
there exists a Weyl reflection $w_\sig \in W_G$ that satisfies 
\ceqn{
\lam_2^{(e,\rbr{\sig})}(H_{\bm{t}} + L_H^c) = w_{\rbr{\sig}}(H_{\bm{t}}) + L_H^c
} 
and the automorphism group $\mathcal{G}_2$ is a subgroup of $W_G$. 
}
\end{coro}
The proof of \refCO{cor5} is given in \refA{proof2}. 
For all the examples studied so far, 
$\Gg$ is $\uni, \su(2), \so(3)$, or their direct sum \cite{Volovik77, Mermin78, Leonhardt00, Kobayashi12a, Schwarz82b}. 
Therefore, it follows from \refCO{cor5} that $\mathcal{G}_2$ is either trivial or a direct sum of $\mathbb{Z}_2$, 
where a possible form of nontrivial \tinf \ is essentially the sign change of a \tcha. 
Since a larger group, in general, has a larger Weyl group, 
it is natural to ask whether other forms appear when we consider a group larger than $G = \mr{SU}(2)$ or SO(3). 
We answer this affirmatively in \refS{6} 
by showing an example of $\mathcal{G}_2 \simeq S_3$.

Finally, there is no topological influence for the case of a three-dimensional skyrmion 
as stated in the following corollary. 
\begin{coro}
\labTH{cor6}\textit{
The \tinf \ on a \thsky \ is trivial. 
}
\end{coro}
\textit{Proof}\\ 
From \refCO{cor3}, we have $\pthgh \simeq \COK i_3^\ast$. 
Then, \refCO{cor6} follows directly from \refTH{theo2}.

\section{\labS{6} Non-Abelian topological influence on a skyrmion in an SU(3)-Heisenberg model}
Since \tinf \ on a monopole and that on a skyrmion are the same 
in that they are characterized by the action \Eref{5.1.1} of $\po$ on $\pt$ \cite{Kobayashi12a}, 
we consider \tinf \ on a skyrmion in the two-dimensional space. 
We include in $G$ and $H$ the space symmetry and the lattice symmetry, respectively, 
because dislocations and disclinations, which result from the breaking of the space symmetry, play a vital role in the \tinf \ analyzed below.

\subsection{\labS{6.A} Vortices and skyrmions in a 3-CDW state}

\subsubsection{\labS{6.A.1} SU(3)-Heisenberg model and its ground state}
The Hamiltonian of the SU(3)-Heisenberg model on a triangular lattice $L$ is given by 
\ceqn{
H = J \sum_{\langle i,j \rangle} \sum_{a=1}^8 T_{a,i} T_{a,j} 
\labE{su3point}
}
where $\langle i,j \rangle$ denotes a pair of nearest-neighbor sites $i$ and $j$, 
and $\tbr{T_{a,i}}_{a=1}^8$ is a set of the generators of $\su(3)$ at site $i$. 
On each site, there are three degenerate states, 
which we refer to as red, green, and blue, and write them as 
\peqn{
\Rs = (1,0,0)^T, \ 
\Gs = (0,1,0)^T, \ 
\Bs = (0,0,1)^T
\labE{threestates}
}
We call these internal degrees of freedom as color. 
This model is expected to be realized in an ultracold atomic gas of alkaline-earth atoms in an optical lattice \cite{Fukuhara07, Desalvo10, Cazalilla09, Gorshkov10} 
and can be regarded as a spin-1 bilinear-biquadratic model with equal bilinear and biquadratic couplings \cite{Papanicolaou88, Lauchli06}. 
For the case of an antiferromagnetic interaction ($J < 0$), 
the ground state $| \Psi \rangle_{\mr{GS}}$ is described by the three-sublattice ordering 
with a periodic alignment of three colors \cite{Tsunetsugu06, Lauchli06, Toth10}, 
and known as the 3-color density-wave state (3-CDW state) \cite{Toth10, Bauer12} 
[see \refF{CDWstate}]: 
\ceqn{
| \Psi \rangle_{\mr{GS}} = \bigotimes_{i \in L_R} \Rs_i \otimes \bigotimes_{i \in L_G} \Gs_i \otimes \bigotimes_{i \in L_B} \Bs_i
\labE{groundstate}
}
where $L_R$, $L_G$, and $L_B$ denote the three sublattices. For the triangular lattice, they are given by 
\ceqn{
&&L_R := \tbr{(m_1 - m_2) \bm{a}_1 + (m_1 + 2 m_2) \bm{a}_2 | m_1, m_2 \in \mbZ}
, \nnco
\\
&&L_G := \tbr{\bm{x} + \bm{a}_2 | \bm{x} \in L_R}
\eco
&&L_B := \tbr{\bm{x} + \bm{a}_1 | \bm{x} \in L_R}
}
where $\bm{a}_1 = (1,0)^T$ and $\bm{a}_2 = (1/2, \sqrt{3}/2)^T$ are the primitive vectors of the triangular lattice in units of the lattice constant $a=1$. 
The 3-CDW state appears as the ground state of the SU(3)-Heisenberg model on various lattices 
including triangular, square, and cubic lattices \cite{Tsunetsugu06, Lauchli06, Toth10, Honerkamp04, Rapp11, Sotnikov14}.

\begin{figure}
\centering
\includegraphics[width=6cm, bb=0 0 360 360]{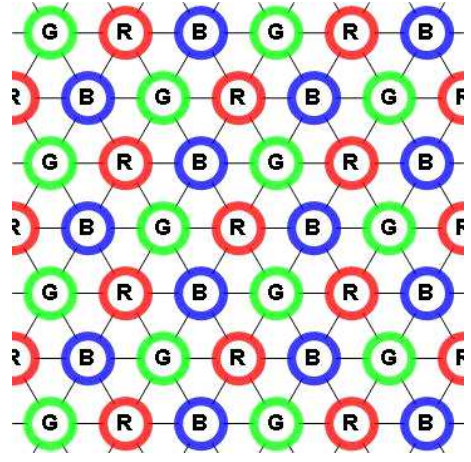}
\caption{\labF{CDWstate} 
(Color online) 
Schematic illustration of the 3-color density-wave state on a triangular lattice. 
The red (R), green (G), and blue (B) disks show the internal states $\Rs, \Gs$, and $\Bs$ defined in \refE{threestates}, respectively. 
The sites in states $\Rs$, $\Gs$, and $\Bs$ constitute the sublattices $L_R$, $L_G$, and $L_B$, respectively. 
}
\end{figure}

\subsubsection{\labS{6.A.2} Symmetries of the system and the state}
When we include the space symmetry, 
the symmetry of the system is given by 
\ceqn{
G = \mr{SU}(3) \times \mr{E}(2)
}
where $\mr{E}(2) := \mbR^2 \rtimes \mr{SO}(2)$ is the two-dimensional Euclidian group 
generated by the two-dimensional translation group $\mbR^2$ and 
the two-dimensional rotational group SO(2), 
where the semidirect product on $H \rtimes N$ is defined by 
$(h,n) \rtimes (h\pri, n\pri) := (h n h\pri n^{-1}, n n\pri)$. 
The ground state $| \Psi \rangle_{\mr{GS}}$ has the continuous symmetry $H_0$ 
generated by diagonal matrices: 
\peqn{
H_0 = \tbr{\exp\br{i \left.\sum_{a=\mr{RG,GB,BR}}c_a H_{\Bal_a^c} } \right| c_a \in \mbR}
\labE{CDWcontinuousH0}
}
Also, $| \Psi \rangle_{\mr{GS}}$ has the discrete symmetries 
that exchange the three colors R, G, and B and three sublattices $L_R, L_G$, and $L_B$ simultaneously. 
The permutations of the colors are described by the symmetry group $S_3$: 
\ceqn{
S_3 &\simeq& \mr{Gen}\tbr{\sig_{RG}, \sig_{GB}, \sig_{BR}}
\nnco
 &\simeq& \tbr{I_3, \sig_{RG}, \sig_{GB}, \sig_{BR}, \sig_{RG}\sig_{GB}, \sig_{GB}\sig_{BR}}
\labE{vortexgroup}
}
where $I_3$ is the identity matrix with size three 
and the generators $\sig_{RG} := \mr{e}^{i \pi S_{RG,1}}, \sig_{GB}:= \mr{e}^{i \pi S_{GB,1}}$, and $\sig_{BR} := \mr{e}^{i \pi S_{BR,1}}$ are given by 
\peqn{
&&\sig_{RG} = \thmat{0}{1}{0}{1}{0}{0}{0}{0}{1}, \ 
\sig_{GB} = \thmat{1}{0}{0}{0}{0}{1}{0}{1}{0}, 
\nnco
&&\sig_{BR} = \thmat{0}{0}{1}{0}{1}{0}{1}{0}{0}
\labE{permutationmatrices}
} 
The permutations of the sublattices are described by the symmetry $H_{\mr{lat}}$ of the lattice: 
\ceqn{
H_{\mr{lat}} &=& \{(m_1\bm{a}_1 + m_2\bm{a}_2, R(n\pi/3)) | 
\nnco
&& \ \ m_1, m_2 \in \mbZ, n = 0,1, \cdots 5 \}
\labE{latticesymm}
}
where $m_1\bm{a}_1 + m_2\bm{a}_2$ and $R(\phi)$ describe translation and rotation, respectively. 
The discrete symmetry is isomorphic to the lattice symmetry: $\dis \simeq H_{\mr{lat}}$. 
Since $h \ (\in H_{\mr{lat}})$ induces a color exchange, 
we define $\sig_h$ as the corresponding matrix in $S_3$. 
From the above discussion, $H$ is generated by the continuous symmetry $H_0$ and the discrete symmetry $H_{\mr{lat}}$: 
\ceqn{
H \simeq H_0 \rtimes H_{\mr{lat}}
}
where $H_0 \rtimes H_{\mr{lat}}$ is the semidirect product defined by $(h_0, h) \rtimes (h_0\pri, h\pri) := (h_0 \ \sig_h h_0\pri \br{\sig_h}^{-1}, h h\pri)$.

\subsubsection{\labS{6.B.1} Vortices in a 3-CDW state}
Vortices and skyrmions in the 3-CDW state are determined in Ref.~\cite{Ueda16}, 
where E(2) and its symmetry breaking are not considered. 
Here we show that the vortices characterized by $S_3$ indeed emerge. 
From \refE{hom1}, we have $\pi_1(G) \simeq \mbZ$, and $\dis \simeq H_{\mr{lat}}$. 
Therefore, we obtain $\mr{Im} \ i_1^\ast \simeq 0$ and $\COKo \simeq \mbZ$ 
and hence  
\peqn{
\po \simeq \mbZ \times_f H_{\mr{lat}}
}
For a \tcha \ $(m, h)$, 
we refer to $\sig_h$ defined above as a spin topological charge. 
For the analysis of \tinf, only a spin \tcha \ is necessary for the two reasons. 
First, vortices described by $\mbZ$ cannot have nontrivial \tinf \ from \refTH{theo2}. 
Second, since skyrmions are shown to be described by an SU(3)-spin texture [see Eqs.~\Eref{RGskyrmion1}, \Eref{RGskyrmion2}, and \Eref{RGskyrmion3}], 
\tinf \ is solely determined by the color exchange $\sig_h$. 
When we focus on spin \tcha s, vortices are characterized by $S_3$: 
\peqn{
\tbr{\sig_h \in S_3 | (m, h) \in \po} \simeq S_3
}
We refer to vortices with spin \tcha s $\sig_{RG}$, $\sig_{GB}$, and $\sig_{BR}$ 
as RG-, GB-, and BR-vortices, respectively, 
according to their exchanges of the colors and sublattices. 
An example of an RG-vortex is the disclination with the Frank angle $\pi/3$ 
around a site belonging to the sublattice $L_B$, 
around which colors R and G and sublattices $L_R$ and $L_G$ are exchanged simultaneously. 
Here, the Frank angle is the angle over which the lattice sites are missing [see Fig.~\Fref{RGvortex} (a) and (d)]. 
Let $\phi$ be the azimuth angle around the vortex. 
Due to the exchange of the sublattices, 
there is an ambiguity in the correspondence between a site and the sublattice it belongs to. 
We therefore fix the range of $\phi$ to $[0, 2\pi)$ to assign one sublattice to each site. 
We take the reference order parameter as the expectation value of $\bm{S}_{\mr{RG}}$ with respect to $| \Psi \rangle_{\mr{GS}}$ in \refE{groundstate}: 
\ceqn{
&&\Ex{\bm{S}_{\mr{RG}}}{R,0} = (0,0,1), \ 
\Ex{\bm{S}_{\mr{RG}}}{G,0} = (0,0,-1), 
\nnco
&&\Ex{\bm{S}_{\mr{RG}}}{B,0} = (0,0,0)
\labE{refOP}
}
where $\Ex{A}{X}$ stands for the expectation value of $A$ over sites of sublattice $L_X$ 
and the subscript $0$ indicates the expectation value with respect to $| \Psi \rangle_{\mr{GS}}$. 
Then, the texture of a vortex is obtained by operating $\exp\br{ i \phi S_{\mr{RG},1} /2}$ 
on the reference order parameter \Eref{refOP}: 
\peqn{
\Ex{\bm{S}_{\mr{RG}}}{R}(\phi) &=& \Ex{ \mr{e}^{ -i {\phi \over 2} S_{\mr{RG},1} } \bm{S}_{\mr{RG}} \mr{e}^{ i {\phi \over 2} S_{\mr{RG},1} }}{R,0}
\nnco
 &=& \rbr{0, \sin\br{\phi \over 2}, \cos\br{\phi \over 2}}
\labE{tex.RGvortex.1}
\eco
\Ex{\bm{S}_{\mr{RG}}}{G}(\phi) &=& \Ex{ \mr{e}^{ -i {\phi \over 2} S_{\mr{RG},1} } \bm{S}_{\mr{RG}} \mr{e}^{ i {\phi \over 2} S_{\mr{RG},1} }}{G,0}
\nnco
 &=& - \rbr{0, \sin\br{\phi \over 2}, \cos\br{\phi \over 2}}
\labE{tex.RGvortex.2}
\eco
\Ex{\bm{S}_{\mr{RG}}}{B}(\phi) &=& \Ex{ \mr{e}^{ -i {\phi \over 2} S_{\mr{RG},1} } \bm{S}_{\mr{RG}} \mr{e}^{ i {\phi \over 2} S_{\mr{RG},1} }}{B,0}
\nnco
 &=& (0,0,0)
\labE{tex.RGvortex.3}
}
One can see from Eqs.~\Eref{tex.RGvortex.1} and \Eref{tex.RGvortex.2} 
that $\bm{S}_{\mr{RG}}$ rotates by angle $\pi$ around the vortex. 
We note that Eqs.~\Eref{tex.RGvortex.1} and \Eref{tex.RGvortex.2} indeed give a continuous map to the OPM because the sublattices $L_R$ and $L_G$ are exchanged at $\phi = 0$ and $2\pi$.

\begin{figure}
\centering
\includegraphics[width=8cm, bb=0 0 921 621]{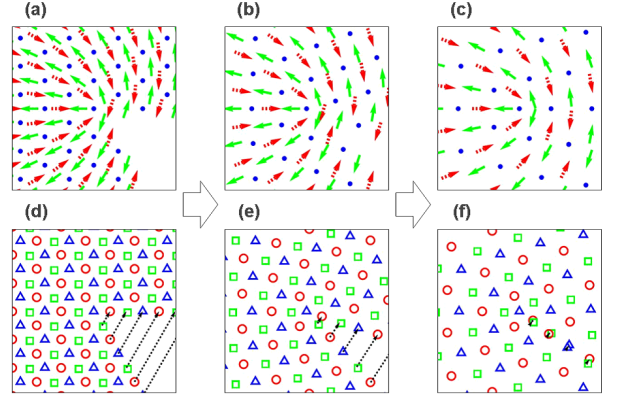}
\caption{\labF{RGvortex} 
(Color online) 
Schematic illustrations of an RG vortex in the 3-CDW state for a triangular lattice. 
It is a disclination with the Frank angle $\pi/3$ 
around a site belonging to the sublattice $L_B$. 
Here, the Frank angle describes the angle over which the lattice sites are missing. 
The texture of the RG-vortex described in Eqs.~\Eref{tex.RGvortex.1}, \Eref{tex.RGvortex.2}, and \Eref{tex.RGvortex.3} is shown in Fig.~(c) 
and Figs.~(a) and (b) illustrate how the RG-spin vortex is obtained as the Frank angle vanishes. 
In the upper panels (a), (b), and (c), 
each arrow indicates the expectation value of $(S_{RG,2}, S_{RG,3})$, 
where the red dashed (green) arrows correspond to the sublattice $L_R$ ($L_G$). 
In the lower panels (d), (e), and (f), 
the red circle, green square, and blue triangule at each site indicate the sublattices $L_R$, $L_G$, and $L_B$, respectively. 
Across the disclination, the sublattices $L_R$ and $L_G$ are exchanged. 
}
\end{figure}

\subsubsection{\labS{6.B.2} Skyrmions in a 3-CDW state}
Since $H$ does not include $\su(2)$-subalgebras from \refE{CDWcontinuousH0}, 
$L_H^c$ vanishes. 
Hence, from \refCO{cor2} 
the second homotopy group is isomorphic to the triangular lattice, which, in turn, is isomorphic to the co-root lattice of SU(3): 
\peqn{
&&\pt \simeq (L_H \cap L_G^c)/L_H^c \simeq L_H \cap L_G^c
\nnco
&\simeq& \tbr{\left. \sum_{a=\mr{RG,GB,BR}} m_a \Bal_a^c \right| m_a \in \mbZ, \sum_{a=\mr{RG,GB,BR}} \Bal_a^c = 0} 
\nnco
&\simeq& L_{\mr{SU}(3)}^c
\labE{3CDWpt}
}
Reflecting the triangular geometry of $L_{\mr{SU}(3)}^c$ [see \refF{corootlattice}], 
the 3-CDW state has three types of skyrmions [see \refF{skyrmions} (c)]. 
Let $(r,\phi)$ be the polar coordinates in $\mbR^2$, 
and $\theta(r)$ be a real function that satisfies $\theta(0) =0$ and $\theta(\infty) = \pi$. 
From \refCO{cor2}, the texture of a skyrmion with \tcha \ $\Bal_{RG}^c$ is obtained by 
operating $g_{\Bal_{RG}}^{(2)}(\theta(r), \phi)$ on the reference order parameter in \refE{refOP}: 
\peqn{
&&\Ex{\bm{S}_{\mr{RG}}}{R}(\theta,\phi) 
\nnco
&=& 
\Ex{ \rbr{g_{\Bal_{RG}}^{(2)}(\theta(r), \phi)}^\dagger \bm{S}_{\mr{RG}} g_{\Bal_{RG}}^{(2)}(\theta(r), \phi) }{R,0}
\nnco
 &=& \hat{\bm{r}}(\theta(r),\phi)
\labE{RGskyrmion1}
\eco
&&\Ex{\bm{S}_{\mr{RG}}}{G}(\theta,\phi) 
\nnco
&=& 
\Ex{ \rbr{g_{\Bal_{RG}}^{(2)}(\theta(r), \phi)}^\dagger \bm{S}_{\mr{RG}} g_{\Bal_{RG}}^{(2)}(\theta(r), \phi) }{G,0}
\nnco
 &=& - \hat{\bm{r}}(\theta(r),\phi)
\labE{RGskyrmion2}
\eco
&&\Ex{\bm{S}_{\mr{RG}}}{B}(\theta,\phi) 
\nnco
&=& 
\Ex{ \rbr{g_{\Bal_{RG}}^{(2)}(\theta(r), \phi)}^\dagger \bm{S}_{\mr{RG}} g_{\Bal_{RG}}^{(2)}(\theta(r), \phi) }{B,0}
\nnco
 &=& 0
\labE{RGskyrmion3}
}
Thus, this skyrmion is described by a hedgehog configuration of $\bm{S}_{\mr{RG}}$ with winding number $+1$ ($-1$) on $L_R$ ($L_G$) as shown in \refF{RGskyrmion}. 
We refer to this skyrmion as an RG-skyrmion. 
Similarly, the texture of a skyrmion with \tcha \ $\Bal_{GB}^c$ is described by a hedgehog configuration of $\bm{S}_{\mr{GB}}$ 
with winding number $+1$ ($-1$) on $L_G$ ($L_B$) 
and we refer to it as a GB-skyrmion  [see \refF{skyrmions} (b)].  
Also, there exists a skyrmion with \tcha \ $\Bal_{BR}^c$ described by a hedgehog configuration of $\bm{S}_{\mr{BR}}$ 
with winding number $+1$ ($-1$) on the sublattice $L_B$ ($L_R$), 
and we refer to it as a BR-skyrmion [see \refF{skyrmions} (c)]. 
It follows from the relation $\Bal_{RG}^c + \Bal_{GB}^c + \Bal_{BR}^c = 0$ [see \refF{corootlattice}] 
that these three skyrmions are not independent; 
the composition of all of them results in a trivial texture.

\begin{figure}
\centering
\includegraphics[width=8cm, bb=0 0 844 583]{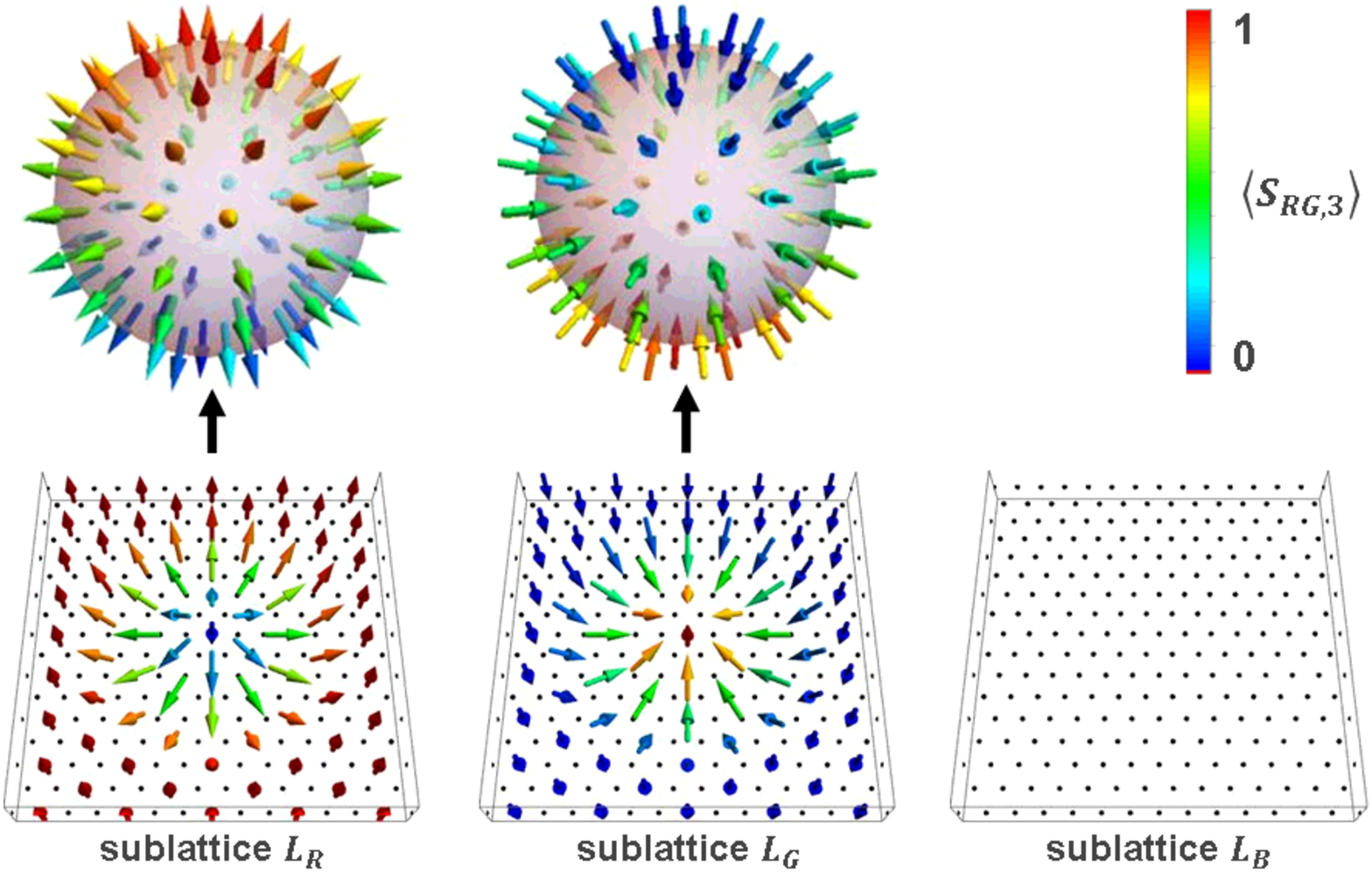}
\caption{\labF{RGskyrmion} 
(Color online) 
Schematic illustrations of an RG skyrmion. 
Each arrow shows the expectation value of the generalized $\su(2)$-spin vector $\Ex{\bm{S}_{\mr{RG}}}{}(\theta,\phi)$, 
where the color represents the third component $\Ex{\bm{S}_{\mr{RG},3}}{}(\theta,\phi)$. 
The absence of an arrow on a site indicates that the expectation value vanishes there. 
When the two-dimensional plane is compactified into a sphere, 
in which the points at infinity are mapped onto the north pole, 
the texture on the sublattice $L_R$ ($L_G$) describes a hedgehog texture of $\Ex{\bm{S}_{\mr{RG}}}{}$ 
with winding number $+1$ ($-1$). 
}
\end{figure}

\begin{figure}
\centering
\includegraphics[width=8cm, bb=0 0 949 728]{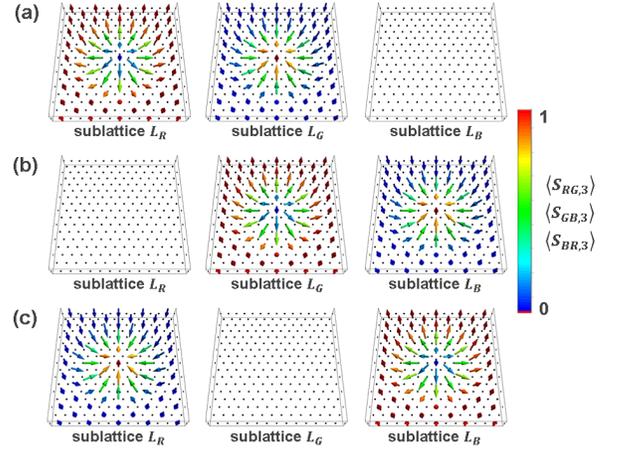}
\caption{\labF{skyrmions} 
(Color online) 
Three types of skyrmions in a 3-CDW state. 
(a) An RG-skyrmion has topological charge $\Bal = \Bal_{RG}^c$ 
and a hedgehog texture of $\bm{S}_{RG}$ with winding number $+1$ ($-1$) on the sublattice $L_R$ ($L_G$).  
(b) A GB-skyrmion has topological charge $\Bal = \Bal_{GB}^c$ 
and a hedgehog texture of $\bm{S}_{GB}$ with winding number $+1$ ($-1$) on the sublattice $L_G$ ($L_B$). 
(c) A  BR-skyrmion has topological charge $\Bal = \Bal_{BR}^c$ 
and a hedgehog texture of $\bm{S}_{BR}$ with winding number $+1$ ($-1$) on the sublattice $L_B$ ($L_R$). 
}
\end{figure}

\subsection{\labS{6.C} Topological influence in a 3-CDW state}
From \refE{3CDWpt} and \refF{corootlattice}, 
$\pt$ is isomorphic to the triangular lattice and $\po$ is isomorphic to $S_3$ 
as far as the spin \tcha \ is considered. 
We will see below that $\mathcal{G}_2$ defined in \refE{autogroup2} is isomorphic to $S_3$, 
where three skyrmions with $\Bal_{RG}, \Bal_{GB}$, and $\Bal_{BR}$ together with their anti-skyrmions with $-\Bal_{RG}, -\Bal_{GB}$, and $-\Bal_{BR}$ 
are exchanged through \tinf, 
reflecting the $S_3$-symmetry of the triangular lattice. 
From \refTH{theo2}, 
the \tinf \ of a vortex with spin \tcha \ $\sig$ on a skyrmion with \tcha \ $n$ is described by the conjugation by $\sig$: 
\peqn{
\lam_2^{\sig}(n) := \sig^{-1} n \sig
}
For example, for $\sig = \sig_{RG}$ and $n = \Bal_{RG}, \Bal_{GB}$, and $\Bal_{BR}$, 
Direct calculations of the matrices in Eqs.~\Eref{su3generators} and \Eref{permutationmatrices} give 
\peqn{
\lam_2^{\sig}(\Bal_{RG}) &=& - \Bal_{RG}, \ 
\lam_2^{\sig}(\Bal_{GB}) = - \Bal_{BR}, 
\nnco
\lam_2^{\sig}(\Bal_{BR}) &=& - \Bal_{GB}
}
A crucial observation here is that this vortex acts on the triangular lattice, 
inverting it about the line perpendicular to $\Bal_{RG}$ [see Fig.~\Fref{CDWinfluence} (a) and (d)]. 
Similarly, a vortex with \tcha \ $\sig_{GB}$ ($\sig_{BR}$) acts on the triangular lattice inverting it about the line perpendicular to $\Bal_{GB}$ ($\Bal_{RG}$) 
as shown in Figs.~\Fref{CDWinfluence} (b) and (e) ((c) and (f)). 
Since $S_3$ is generated by $\sig_{RG}$, $\sig_{GB}$, and $\sig_{BR}$, 
we have $\mathcal{G}_2 \simeq S_3$.

\begin{figure}
\centering
\includegraphics[width=8cm, bb=0 0 943 669]{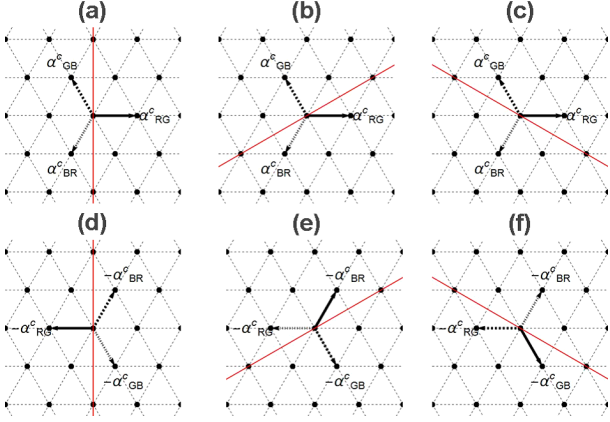}
\caption{\labF{CDWinfluence} 
(Color online) 
Topological charges of skyrmions (a), (b), and (c) before and (d), (e), and (f) after making a complete circuit of the RG-, GB-, and BR-vortices, respectively. 
The RG-, GB-, and GR-vortices act on skyrmions by inverting them about 
the line perpendicular to $\Bal_{RG}$, $\Bal_{GB}$, and $\Bal_{BR}$, respectively. 
}
\end{figure}

The non-Abelian property of $\mathcal{G}_2$ emerges when we consider \tinf \ of two vortices. 
Let $\sig$ and $\tau$ be the \tcha s of the vortices and $n$ be that of a skyrmion. 
Suppose that the skyrmion goes around the vortex with $\sig$ clockwise, 
goes around the vortex with $\tau$ clockwise, 
goes around the vortex with $\sig$ anticlockwise, 
and finally goes around the vortex with $\tau$ anticlockwise [see \refF{2vor1sky}]. 
Since the third (fourth) process is the inverse process of the first (second) one, 
the change in the \tcha \ is given by 
\neqn{
\lam_2^{\tau^{-1}}\br{\lam_2^{\sig^{-1}}\tbr{\lam_2^{\tau}\rbr{\lam_2^{\sig}(n)}}} =  \lam_2^{\rho}(n) 
\FOR \rho = \tau^{-1}  \sig^{-1} \tau \sig
. \nnco
}
While the final topological charge coincides with the initial one for an Abelian $\mathcal{G}_2$ because we have $\rho= e$ for any pair of vortices, 
it does not for a non-Abelian $\mathcal{G}_2$ because $\rho \neq e$ in general. 
For the case of $\sig = \sig_{RG}, \tau = \sig_{GB}$, and $n = \Bal_{RG}$, 
we have $\rho = \sig_{BR}\sig_{GB}$ and $\lam_2^{\rho} (\Bal_{RG})  = \Bal_{GB} \neq \Bal_{RG}$.

\begin{figure}
\centering
\includegraphics[width=8cm, bb=0 0 932 452]{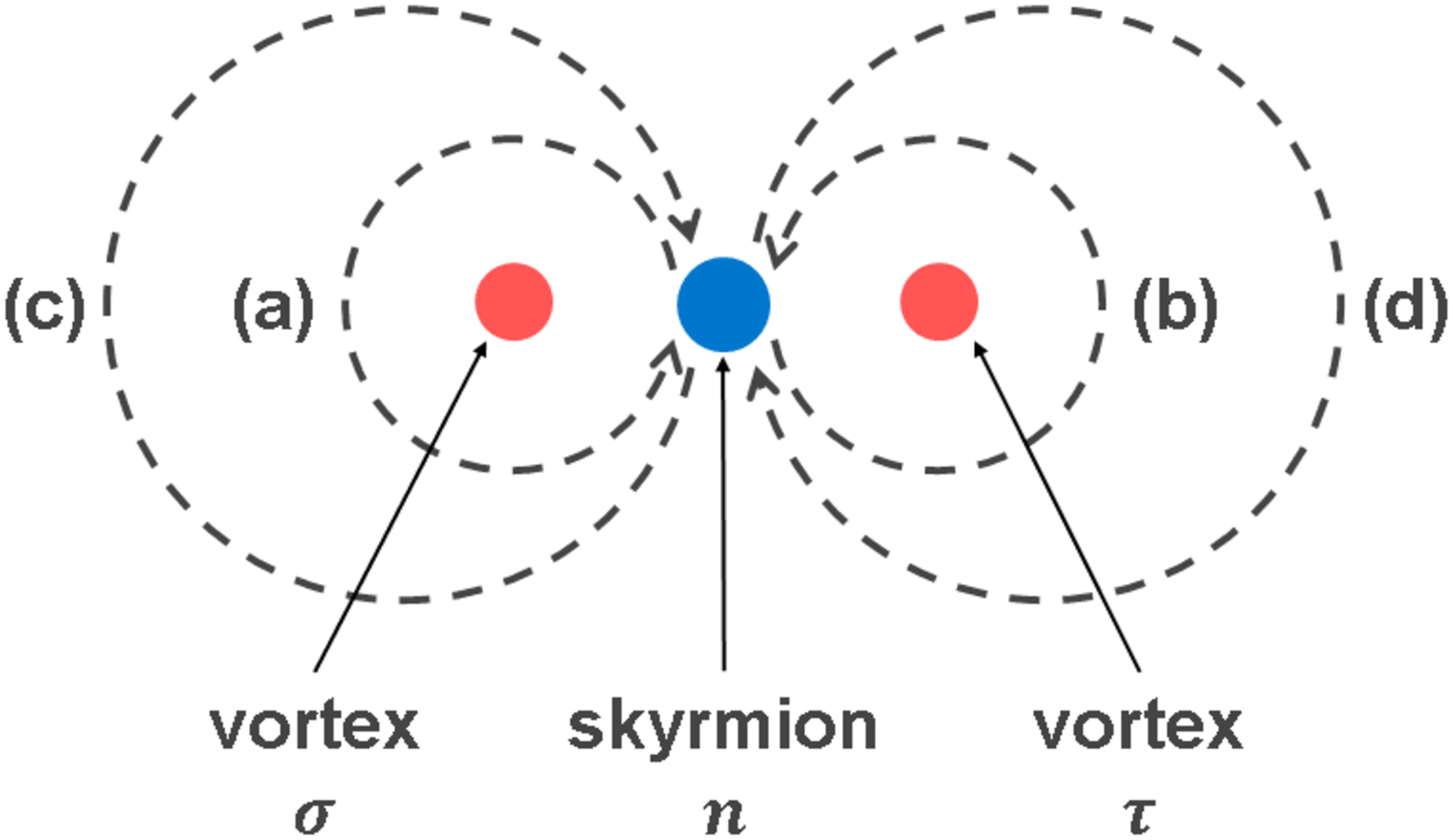}
\caption{\labF{2vor1sky} 
(Color online) 
Topological influence of two vortices with spin \tcha s $\sig$ and $\tau$ on a skyrmion with \tcha \ $n$. 
The skyrmion goes around the vortex with $\sig$ clockwise (a), 
then around the vortex with $\tau$ clockwise (b), 
then around the vortex with $\sig$ anticlockwise (c), 
and finally around the vortex with $\tau$ anticlockwise (d). 
Through these processes, 
the \tcha \ of a skyrmion changes from $n$ to $\lam_2^{\rho}(n)$ 
with $\rho = \tau^{-1}  \sig^{-1} \tau \sig$. 
}
\end{figure}


\section{\labS{conclusion}Conclusion and discussion}
In the present paper, 
we have developed a general method to determine the homotopy group $\pim$ of the order parameter manifold $G/H$ 
by deriving the formula \Eref{homm} which expresses $\pim$ in terms of $\pi_m(G)$ and $\pi_m(H)$. 
Since the homotopy group of a Lie group and each texture on it can be calculated systematically 
by means of the Cartan canonical forms \Eref{Cartanform} and the lattices defined in  Eqs.~\Eref{integrallattice} and \Eref{corootlattice}, 
the obtained formulas allow us to calculate $\pim$ and the texture $O$ of each \te \ systematically. 
We find that 
the textures of a monopole and that of a \thsky \ are obtained by the replacement of the $\su(2)$-spin vector $\bm{S}$ by the generalized $\su(2)$-spin vector $\bm{S}_{\Bal}$ defined in \refE{GM}, 
and that their \tcha s are described by a set of co-roots, 
reflecting the fact that a Lie algebra $\Gg$, in general, includes multiple $\su(2)$-subalgebras. 
We have also shown the necessity of a discrete symmetry $\dis$ for the presence of nontrivial \tinf. 
Moreover, we derive the necessary and sufficient condition for the presence of non-Abelian vortices 
and prove the absence of \tinf \ on a \thsky. 
As for \tinf \ on a \mon \ or a skyrmion, 
we prove that the automorphism group $\mathcal{G}_2$ of topological influence is a subgroup of the Weyl group $W_G$, 
clarifying why only one type of topological influence is known so far. 
Seeking for other types, we find that \tinf \ characterized by a non-Abelian group $S_3$ 
emerges in the 3-color density-wave state of the SU(3)-Heigenberg model, 
where three types of skyrmions and vortices characterized by $S_3$ appear. 
These skyrmions change their types through the \tinf, 
giving $\mathcal{G}_2 = S_3$. 

Finally, we raise three problems for future study. 
First, the dynamical stability of the textures of \te s derived in \refS{3} needs to be clarified. 
These textures and their variations have widely been used as candidates for dynamically stable textures of \te s 
\cite{Nakahara03, Nagaosa13, Vollhardt13, Volovik03, Kawaguchi12, Manton04}. 
In fact, the dynamically stability has been demonstrated in a number of examples 
\cite{Manton04, Tong05, Sondhi93, Yang06, Roszler06, Sampaio13, Ueda16}. 
The texture $O(\theta, \phi) := g_{\Bal}^{(2)}(\theta,\phi) O_0$ and its variation $O(r, \phi) := g_{\Bal}^{(2)}[\theta(r),\phi] O_0$ are widely used 
as candidates for the textures of a monopole and that of a skyrmion, respectively \cite{Manton04, Kawaguchi12, Vollhardt13, Volovik03, Nagaosa13, Nakahara03}, 
where $\theta(r)$ is a function subject to the boundary conditions $\theta(0)=0$ and $\theta(\infty)=\pi$. 
Moreover, they indeed give stable textures \cite{Pietila09, Tong05, Manton04, Sondhi93, Yang06, Roszler06, Sampaio13, Ueda16} 
for an appropriate choice of $\theta(r)$. 
It merits further study to clarify their dynamical stability. 
Second, we represent in \refCO{cor4} the necessary and sufficient condition for non-Abelian vortices 
in terms of the map $f$ defined in \refE{factorsets1}. 
However, its physical implication is yet to be clarified. 
Considering the growing interest in the dynamics of non-Abelian vortices \cite{Kobayashi09, Kobayashi16, Priezjev02, Mcgraw98, Spergel97}, 
it is of interest to understand whether we can simplify the condition \Eref{nonAbelianvortex}. 
Third, 
analogous concepts of topological influence in topological insulators and superconductors 
have recently been discussed in specific examples \cite{Moore08, Kennedy15, Kennedy16}, 
where the domain $S^m$ of $\pim$ and the order parameter manifold $G/H$ are 
replaced by a Brillouin zone and a space of Hamiltonians, respectively. 
When a lower-dimensional topological invariant is nontrivial, 
one can change a higher-dimensional topological invariant under continuous deformation of a Hamiltonian. 
It is worthwhile to analyze general conditions for nontrivial topological influence in topological insulators and superconductors 
and clarify its difference from topological influence in topological excitations 
by using the general formulas developed in the present paper.

\begin{acknowledgments}
S. H. thanks Y. Akagi and N. Kura for helpful discussions. 
This work was supported by
KAKENHI Grant No. JP26287088 from the Japan Society for the Promotion of Science, 
a Grant-in-Aid for Scientific Research on Innovative Areas ``Topological Materials Science'' (KAKENHI Grant No. JP15H05855), 
and the Photon Frontier Network Program from MEXT of Japan. 
S. H. acknowledges support from the Japan Society for the Promotion of Science through the Program for Leading Graduate Schools (ALPS) and JSPS fellowship (JSPS KAKENHI Grant No. JP16J03619). 
\end{acknowledgments}

\appendix

\section{\labS{proof1}Proof of \refLM{lemm1}}
\refLM{lemm1} follows from 
the following theorem on the third homotopy group of a simple compact Lie group \cite{Onishchik94}. 
\begin{theo}
\labTH{theo3}\textit{
Let $G$ and $\bm{S}_{\Bal}$ be a simple compact Lie group and 
the generalized $\su(2)$-spin vector for the co-root $\Bal^c$ with shortest length in $G$, respectively. 
Then, we have 
\ceqn{
\pi_3(G) \simeq \tbr{\left. m \rbr{g_{\Bal}^{(3)}}_G \right| m \in \mathbb{Z}}
\labE{lemm1eq}
}
where $\rbr{g_{\Bal}^{(3)}}_G$ denotes the homotopy class of $G$ with representative element $g_{\Bal}^{(3)}$ defined in \refErm{hom3.02}. 
The isomorphism in \refErm{lemm1eq} is given by $i_{\ast 3}: \pi_3(H) \to \pthgh$, 
where $H\pri = \mr{SU}(2)$ or $\mr{SO}(3)$ is the subgroup of $G$ generated by $\bm{S}_{\Bal}$. 
The right-hand side of \refErm{lemm1eq} does not depend of the choice of the co-root 
since $\rbr{g_{\Bal_1}^{(3)}}_G = \rbr{g_{\Bal_2}^{(3)}}_G$ for two co-roots $\Bal_1^c$ and $\Bal_2^c$ with the shortest length. 
}
\end{theo}

\textit{Proof of \refLM{lemm1}}\\
Let \refE{homm3.01} and $\Bal_i^c$ be the decomposition of the Lie algebra $\Gg$ of $G$ and a co-root in $\Gg_i$, respectively. 
If we denote the universal covering group of $G\pri$ by $\widetilde{G\pri}$, 
$\widetilde{G}$ is given by $\mbR^{a\pri} \times \widetilde{G}_1 \times \cdots \times \widetilde{G}_a$ 
and hence we obtain
\ceqn{
\pi_3(G) &\simeq& \pi_3\br{\mbR^{a\pri} \times \widetilde{G}_1 \times \cdots \times \widetilde{G}_a}
 \simeq \bigoplus_{i=1}^a \pi_3(\widetilde{G}_k) 
\nnco
 &\simeq& \tbr{\left. \sum_{i=1}^a m_i \rbr{g_{\Bal_i}^{(3)}} \right| m_i \in \mbZ} 
\labE{prooftheo3.1}
}
where the first isomorphism follows from the relation $\pi_m(G\pri) \simeq \pi_m(\widetilde{G}\pri) \FOR \forall m \ge 2$, 
the second one from the relation $\pi_m(X \times Y) \simeq \pi_m(X) \oplus \pi_m(Y)$, 
and the third one from \refTH{theo3}, 
which completes the proof of \refLM{lemm1}.

\section{\labS{proof2}Proof of \refTH{theo1}}
\refTH{theo1} is proved by applying the theory of a group extension with an Abelian kernel \cite{Robinson96}. 

\subsection{\labS{B.1} Group extension with an Abelian kernel}
\begin{defi}[Group extension]
$($a$)$ 
\textit{
Let $Q$ and $N$ be two groups. 
Then $G$ is a group extension of $Q$ by $N$ 
if $N$ is a normal subgroup of $G$ and $Q$ is a quotient group of $G$ by $N$, i.e., $Q = G/N$. 
In particular, if $N$ is Abelian, $G$ is referred to as a group extension of $Q$ by an Abelian kernel $N$. 
}
\textit{
$($b$)$ 
Let $G$ and $G\pri$ be group extensions of $Q$ by $N$. 
We denote the projection from $G$ $(G\pri)$ to $Q$ as $T$ $(T\pri)$. 
If there exists an isomorphism $F: G \to G\pri$ such that $T\pri \circ F = T$, 
the two group extensions $G$ and $G\pri$ are regarded as equivalent. 
}
\end{defi}
Let $G, N$, and $Q$ be a group, a normal subgroup of $G$, and the quotient group of $G$ by $N$, respectively, 
and consider a situation in which we know $N$ and $Q$ but do not know $G$. 
The problem of constructing $G$ from $N$ and $Q$ is referred to as an group extension problem. 
As we will see below, there is a general theory to solve the group extension problem if $N$ is Abelian. 

Let $T$ be the projection from $G$ to $Q$. 
A map $s : Q \to G$ that satisfies $T \circ s = \mr{id}_Q$ is referred to as a section of $T$, 
where $\mr{id}_Q$ denotes the identity map on $Q$. 
We assume that $N$ is Abelian and that a section $s$ of $T$ is given. 
We define the map $f : Q \times Q \to N$ referred to as the factor set of $G$ associated with $s$ by 
\peqn{
f(q,q\pri) := s(q) s(q\pri) s(qq\pri)^{-1}
\labE{deff}
}
Since $T\rbr{f(q,q\pri)} = e$ and hence $f(q,q\pri) \in N$, $f$ is indeed a map to $N$. 
Since $N$ is a normal subgroup of $G$, 
$N$ is invariant under the inner isomorphism $g\pri \mapsto g g\pri g^{-1}$. 
Moreover, the inner isomorphisms of $N$ acts on $N$ trivially because $N$ is Abelian. 
Therefore, 
the inner isomorphism of $g \in G$ depends only on the quotient element $q = T(g)$. 
We define the map $\theta_q: N \to N$ for $q$ by 
$\theta_q(n) := g n g^{-1}$, where $g \in G$ satisfies $q = T(g)$. 
Then, the following theorem holds \cite{Robinson96}. 
\begin{theo}
\labTH{groupext}
$($a$)$ 
\textit{
Let the product $\times_f$ on the product set $N \times Q$ be defined by 
\ceqn{
(n,q) \times_f (n\pri,q\pri) := (n + \theta_q(n\pri) + f(q,q\pri), qq\pri)
}
where we write the product on $N$ by the sum because $N$ is Abelian. 
Then, $N \times Q$ becomes a group, 
where the identity element is given by $(f(e,e)^{-1},e)$ and the inverse of $(n,q)$ is given by 
$(n^{-1}+ f(q,q^{-1})^{-1}, q^{-1})$. 
This group denoted by $N \times_f Q$ is a group extension of $Q$ by $N$ and isomorphic to $G$ under this product. 
}
$($b$)$
\textit{
Let $\bar{s}$ and $\bar{f}$ be another section of $T$ 
and the factor set associated with $\bar{s}$, respectively. 
If there exists a map $\alpha: Q \to N$ that satisfies $\bar{s}(q) = \alpha(q)s(q)$, 
the two group extensions $N \times_f Q$ and $N \times_{\bar{f}} Q$ are equivalent. 
}
\end{theo}

\subsection{\labS{B.2} Proof of \refTH{theo1}}
We start from the relation derived in Ref.~\cite{Higashikawa16a}: 
\ceqn{
\frac{\pim}{\CIM} \simeq \KIM
\labE{shortexact}
}
which follows from the homotopy exact sequence \cite{Higashikawa16a, Hatcher02}. 
From the homotopy lifting theorem \cite{Hatcher02}, 
any homotopy class $O$ of $G/H$ can be written as 
\ceqn{
O(\bm{x}) = g(\bm{x}) O_0
\FOR \forall \bm{x} \in D^m
\labE{proof.theo1.1}
}
where $g$ is a map from $D^m$ to $G$ 
subject to the boundary condition 
\peqn{
g(\bm{x}) = e
\FOR \norm{\bm{x}} = \pi 
}
Then, the projection map $T: \pim \to \KIM$ in \refE{shortexact} is given by 
\ceqn{
\rbr{T(O)}(\hat{\bm{x}}) := \lim_{r \to 0} g^O(r \hat{\bm{x}}) 
\FOR \hat{\bm{x}} \in S^{m-1}
}
where $T(O)$ is a map from $S^{m-1}$. 

We first prove that the inner isomorphism of $\pim$ on $\CIM$ is trivial: 
\neqn{
n [a] n^{-1} = [a]
\FOR \forall n \in \pim, \forall [a] \in \CIM
. \nnco
\labE{equ1}
}
For $m \ge 2$, \refE{equ1} follows from the commutativity of higher-dimensional homotopy groups. 
For $m = 1$, from \refE{proof.theo1.1}, we can express the loop $l^n$ ($l^{[a]}$) corresponding $n$ ($[a]$) as 
$l^n(\phi) = \gamma_n(\phi) O_0$ ($l^{[a]}(\phi) = a(\phi) O_0$), 
where $\gamma_n$ is a path from $\gamma_n(0) = \sig_n$ to $\gamma_n(2\pi) = e$ 
and $a$ is a loop on $G$. 
Then, defining $a_s(\phi)$ for $s,\phi \in [0,2\pi]$ by 
\neqn{
&& a_s(\phi) 
\nnco
&& :=
\begin{cases}
\gamma_n(2\pi - 3\phi)O_0 & 0 \le \phi \le \frac{s}{3}
; \\
a \br{\frac{\pi (3\phi - s)}{3\pi - s}}\gamma_n(2\pi - s)O_0 & \frac{s}{3} \le \phi \le 2\pi - \frac{s}{3}
; \\
\gamma_n(3\phi - 4\pi)O_0 & 2\pi - \frac{s}{3} \le s \le 2\pi
, 
\end{cases}
\nnco
}
we find that $a_s$ is a continuous deformation from $a_{s=0} = [a]$ to $a_{s=2\pi} = n[a]n^{-1}$, 
which completes the proof of \refE{equ1}. 

We next apply \refTH{groupext} to derive Eqs.~\Eref{homm3}, \Eref{homm}, and \Eref{homm4}. 
Let $b$ and $\rbr{O^b}_{G/H}$ be an element of $\KIM$ and that of $\pim$ defined in \refE{tex.ker}, respectively. 
When we define $S:\KIM \to \pim$ by $S(b) := \rbr{O^b}_{G/H}$, 
it is a section of $T$ 
since 
\peqn{
\tbr{\rbr{T \circ S}(b)}(\bm{x}) &=& \rbr{T(\rbr{O^b}_{G/H})}(\bm{x}) = b_{s=0}\br{ \hat{\bm{x}} } 
\nnco 
&=& b(\bm{x}) \FOR \bm{x} \in S^{m-1}
}
We define $f : \KIM \times \KIM \to \CIM$ by \refE{deff} with substitution of $S$ for $s$. 
Since $\CIM$ is Abelian for any $m$, 
\refTH{groupext} gives the isomorphism: 
\ceqn{
\pim \simeq \CIM \prf \KIM
\labE{proof1}
}
where the product on the right-hand side of \refE{proof1} is defined by 
\neqn{
([a], b) \times_f ([a\pri],b) := ([a] + \theta_b([a\pri]) + f(b,b\pri), bb\pri)
. \nnco
\labE{Ap_prod}
}
From \refE{equ1}, the inner isomorphism of $\pim$ on $\CIM$ is trivial: $\theta_b([a\pri]) :=b[a\pri]b^{-1} = [a\pri]$, 
which gives Eqs.~\Eref{homm3} and \Eref{homm}. 
Since we take the section of the identity element as $S(e) := \rbr{O^e}_{G/H} = e$, we obtain $f(e,b) =S(e) S(b) \rbr{S(e b)}^{-1} = e$ and hence \refE{homm4}. 

We finally prove that any choice of the section gives an equivalent group extension. 
Let $\rbr{O^b}_{G/H}$ be another element of $\pim$ corresponding to another deformation $\bar{b}_s$ of $b \in \KIM$ to the trivial homotopy class. 
Then, the map $\bar{S}: b \mapsto \rbr{\bar{O}^b}_{G/H}$ provides another section. 
From the relation 
\neqn{
&&\rbr{(O^b)^{-1} \bar{O}^b}(\bm{x}) 
\nnco
&=& 
\begin{cases}
b_{s=\pi - 2\norm{x}}\br{ \hat{\bm{x}} } O_0
&\FOR 0 \le \norm{x} \le {\pi \over 2}
; \\
\bar{b}_{s=2\norm{x} - \pi}\br{ \hat{\bm{x}} } O_0
&\FOR {\pi \over 2} \le \norm{x} \le \pi
, 
\end{cases}
}
we have 
\ceqn{
\bar{b}_{s= 2\norm{x} - \pi}\br{ \hat{\bm{x}} } &=& e \FOR \norm{x} = \pi
\eco
\bar{b}_{s= \pi - 2\norm{x} }\br{ \hat{\bm{x}} } &=& e \FOR \norm{x} = 0
}
and hence $(\rbr{O^b}_{G/H})^{-1} \rbr{\bar{O}^b}_{G/H} \in \CIM$. 
Therefore, defining the map $\alpha: \KIM \to \CIM$ by $\alpha(b) := \rbr{S(b)}^{-1} \bar{S}(b)$, 
we find from \refTH{groupext} (b) that the two sections $S$ and $\bar{S}$ give the equivalent group extensions, 
which completes the proof of \refTH{theo1}.

\section{\labS{proofcor2} Proof of \refCO{cor2}}
Since $\ptg$ vanishes from \refE{homG2}, $\COK i_2^\ast$ vanishes. 
It follows from \refTH{theo1} that 
\peqn{
\pt \simeq \mr{Ker}\tbr{ i_1^\ast: \poh \to \pog}
\labE{proofcor2}
}
Since an element of $L_H^c$ describes a trivial loop on $H$, 
it is trivial as a loop on $G$, 
which indicates $L_H^c \subset L_G^c$. 
Since $L_H^c \subset L_H$, 
$L_H^c$ is an Abelian subgroup of $L_H \cap L_H^c$. 
Let us write elements of $\poh \simeq L_H/L_H^c$ as $H_{\bm{t}} + L_H^c$ ($H_{\bm{t}} \in L_H$) and those of $\pog \simeq L_G/L_G^c$ as $H_{\bm{t}} \in L_G$ ($H_{\bm{t}} + L_G^c$). 
For an element $H_{\bm{t}} + L_H^c$ of $\pt \simeq \mr{Ker} \ i_1^\ast$, 
we have $i_1^\ast(H_{\bm{t}} + L_H^c)  = e$, and hence $H_{\bm{t}} + L_H^c \in L_G^c$.  We thus obtain Eqs.~\Eref{hom2} and \Eref{hom22}: 
\peqn{
\pt &\simeq& \tbr{ H_{\bm{t}} + L_H^c | H_{\bm{t}} \in L_H, H_{\bm{t}} + L_H^c \in L_G^c}
\nnco
 &\simeq& \tbr{ H_{\bm{t}} + L_H^c | H_{\bm{t}} \in L_H \cap  L_G^c}
\nnco
 &\simeq& (L_H \cap L_G^c)/L_H^c
}
We next derive the texture \Eref{hom23} for the \tcha \ $n$ given in \refE{hom22}, 
which can be done straightforwardly from the construction \Eref{tex.ker} and \refTH{theo1}. 
The right-hand side of \refE{hom22} describes the loop 
\ceqn{
g_n(\phi) = \exp\br{ i \phi \sum_{j=1}^r m_j H_{\Bal_j^c}} 
\labE{proofcor2.1}
}
which can continuously transform into a trivial one 
through the continuous deformation $g(\theta, \phi)$ defined by 
\neqn{
&&g(\theta, \phi) := \mr{e}^{-i\theta S_{\Bal_1,2}} g_{\Bal_1}^{(2)}(\theta,m_1\phi) \mr{e}^{-i\theta S_{\Bal_2,2}}g_{\Bal_2}^{(2)}(\theta,m_2\phi)
\nnco
&&\times \cdots \times \mr{e}^{-i\theta S_{\Bal_r,2}}g_{\Bal_r}^{(2)}(\theta,m_r\phi) 
\exp\br{i \phi \sum_{j=1}^r m_j S_{\Bal_j,3}}
, \nnco
\labE{proofcor2.2}
}
where $\theta \ (\in [0,\pi])$ is the parameter of the deformation. 
In fact, since 
\ceqn{
g_{\Bal_j,1}^{(2)}(0,m_j\phi) &=& \mr{e}^{i \phi m_j S_{\Bal_j,3}}
\eco
\mr{e}^{-i\pi S_{\Bal_j,2}} g_{\Bal_j}^{(2)}(\pi,m_j\phi)
 &=& \mr{e}^{-i\pi S_{\Bal_j,2}} \mr{e}^{i \phi m_j S_{\Bal_j,3}} \mr{e}^{i\pi S_{\Bal_j,2}} 
\nnco
 &=& \mr{e}^{- i \phi m_j S_{\Bal_j,3}}
}
we obtain $g(\theta = 0, \phi) = g_n(\phi)$ and $g(\theta = \pi, \phi) = e$. 
Thus, the texture $\widetilde{O}(\theta,\phi)$ is given by 
\peqn{
&&\widetilde{O}(\theta,\phi) = \mr{e}^{-i\theta S_{\Bal_1,2}} g_{\Bal_1}^{(2)}(\theta,m_1\phi) \mr{e}^{-i\theta S_{\Bal_2,2}}g_{\Bal_2}^{(2)}(\theta,m_2\phi)
\nnco
&&\times \cdots \times \mr{e}^{-i\theta S_{\Bal_r,2}}g_{\Bal_r}^{(2)}(\theta,m_r\phi)O_0
}
Let us define 
\ceqn{
\widetilde{O}_u(\theta,\phi) = g_{u,1}(\theta,\phi)g_{u,2}(\theta,\phi) 
\cdots g_{u,r}(\theta,\phi) O_0
\labE{defOu}
}
where $g_{u,b}(\theta,\phi) := \mr{e}^{-i\theta(1-u) S_{\Bal_j,2}} g_{\Bal_j}^{(2)}(\theta,m_j\phi)$ and $u \in [0,1]$. 
Equation~\Eref{defOu} gives a continuous deformation from $O_{u=0} = \widetilde{O}$ to $O_{u=1} = O$, 
because $g_{u,b}$ is a continuous deformation 
from $g_{u=0,b}(\theta,\phi) = \mr{e}^{-i\theta S_{\Bal_j,2}} g_{\Bal_j}^{(2)}(\theta,m_j\phi)$ 
to $g_{u=1,b}(\theta,\phi) = g_{\Bal_j}^{(2)}(\theta,m_j\phi)$, 
which completes the proof of \refCO{cor2}.

\section{\labS{prooftheo2} Proof of \refTH{theo2}}
It follows from \refE{homm3} that 
the action of $([H_{\bm{t}}], [\sig])$ can be decomposed into 
the action of $([H_{\bm{t}}], e)$ and that of $(e, [\sig])$ as follows: 
\peqn{
\lam_m^l(n) = \lam_m^{(e, [\sig])}\tbr{\lam_m^{([H_{\bm{t}}], e)}\rbr{([a],b)}}
\labE{Proof_theo2,1}
}
From \refE{hom1.1.1}, we can write the texture $O_1(\phi)$ of a vortex with \tcha \ $([H_{\bm{t}}], e)$ as $O_1(\phi) := g_1(\phi) O_0$, 
where $g_1(\phi) := \exp(i \phi H_{\bm{t}})$ and $\phi \ (\in [0,2\pi])$ is the azimuth angle around the vortex. 
Let $O^{([a],b)}(\bm{x})$ be the texture of a \te \ with \tcha \ $([a],b)$ 
and let us define $\lam_s$ for $s \in [0,2\pi]$ by 
\neqn{
&&\lam_s(\bm{x}) 
\nnco
&&:=
\begin{cases}
g_1(s) O^{([a],b)}\rbr{\br{{\pi \over {\pi\over 2} + {s \over 4}}} \bm{x}}
& \FOR 0 \le \norm{x} \le {\pi\over 2} + {s \over 4}
; \\
g_1(4 \norm{x} - 2 \pi) O_0 
& \FOR {\pi\over 2} + {s \over 4} \le \norm{x} \le \pi
. 
\end{cases}
\nnco
}
Since $g_1(2\pi) = e$, $\lam_s$ is a continuous deformation from $\lam_{s=0}(\bm{x}) = \lam_m^{(\rbr{H_{\bm{t}}}, e)}([a],b)$ to 
\ceqn{
\lam_{s=2\pi}(\bm{x}) =
 O^{([a],b)}\br{ \bm{x} } = O^{([a],b)}\br{ \bm{x} }
}
subject to the boundary condition \Eref{boundarym2}. 
We thus have $\lam_m^{([H_{\bm{t}}], e)}\rbr{([a],b)} = ([a],b)$. 
Then, \refE{Proof_theo2,1} reduces to 
\ceqn{
&&\lam_m^{([H_{\bm{t}}], [\sig])}\rbr{([a],b)} 
\nnco
&=& \lam_m^{(e, [\sig])}\rbr{([a],b)} 
= \lam_m^{(e, [\sig])}\rbr{([a],e) \times_f (e,b)}
\nnco
 &=& \lam_m^{(e, [\sig])}\rbr{([a],e)} \times_f \lam_m^{(e, [\sig])}\rbr{(e,b)}
}
where we use the homomorphic property of $\lam_m^l$ in the third equality. 
Let $\gamma_\sig$ be a path from $\gamma_\sig(0)=\sig$ to $\gamma_\sig(2\pi) = e$ 
and we write the texture of a \te \ with \tcha \ $([a],e)$ ($(e,b)$) by $O\pri(\bm{x}) = g(\bm{x}) O_0$ 
with $g(\bm{x}) = a(\bm{x})$ ($g(\bm{x}) = b_{\norm{x}}( \hat{\bm{x}} )$). 
Let us define $\lam_s\pri$ for $s \in [0,2\pi]$ by 
\neqn{
&&\lam_s\pri(\bm{x})
\nnco
& := &
\begin{cases}
\gamma_\sig(s)\sig^{-1} O\pri\rbr{\br{{\pi \over {\pi\over 2} + {s \over 4}}} \bm{x}}
& \FOR 0 \le \norm{x} \le {\pi\over 2} + {s \over 4}
; \\
\gamma_\sig\br{4 \ \norm{x} - 2 \pi} O_0 
& \FOR {\pi\over 2} + {s \over 4} \le \norm{x} \le \pi
. 
\end{cases}
\nnco
}
Then, $\lam_s\pri$ is a continuous deformation from 
$\lam_{s=0}\pri(\bm{x}) = \lam_m^{(e,[\sig])}(\bm{x})$ to 
\neqn{
\lam_{s=2\pi}\pri(\bm{x}) = \sig^{-1} O\pri(\bm{x}) = \sig^{-1} g(\bm{x}) \sig O_0
 = \rbr{\sig^{-1} g \sig} (\bm{x})
, \nnco
\labE{prooftheo2.1}
}
where we use $\sig \in H \cap G_0$ and hence $\sig O_0 = O_0$ in the second equality in \refE{prooftheo2.1}. 
When the \tcha \ is $([a],e)$, 
$\rbr{\gamma_\sig(s)}^{-1} a \gamma_\sig(s)$ for $s\in [0,2\pi]$ 
describes a continuous deformation from $\sig^{-1} a \sig$ to $a$ 
subject to the boundary condition \Eref{boundarym2}. 
We therefore have $\lam_m^{(e,[\sig])}\rbr{([a],e)} = ([a],e)$ and hence \refE{5.1.1}: 
\ceqn{
&&\lam_m^{(e, [\sig])}\rbr{([a],e)} \times_f \lam_m^{(e, [\sig])}\rbr{(e,b)}
\nnco
 &=& ([a],e) \times_f (e,\sig^{-1} b \sig) 
 = ([a],\sig^{-1} b \sig)
}
which completes the proof of \refTH{theo2}.

\section{\labS{proof3} Proof of \refCO{cor5}}
Let $\Gg_C$ be the Cartan subalgebra of $G$ 
and we define two subalgebras of $\Gg$ by
\ceqn{
\Gh_C^\perp &:=& \tbr{H_{\bm{u}} \in \Gg_C | (\bm{u}, \bm{t}) = 0 \FOR \forall H_{\bm{t}} \in L_H \cap L_G^c}
, \nnco
\labE{defgcfp}
\nnco
\Gh^\perp &:=& \Gh_C^\perp \oplus \SPAN\{E_{\Bal}^R, E_{\Bal}^I | \Bal \in R_+,
\nnco
&&(\Bal, \bm{t}) = 0 \FOR \forall H_{\bm{t}} \in L_H \cap L_G^c\}
\labE{defgfp}
}
where $\mr{Span}S$ denotes the vector space spanned by the elements of $S$. 

We first prove that $\Gh^\perp$ is a subalgebra of $\Gg$ that commutes with $L_H \cap L_G^c$. 
From the commutation relations of the Cartan canonical form, 
the commutators among $H_{\bm{u}} \in \Gh_C^\perp$ and  $E_{\Bal}^{R,I} \in \Gh^\perp$ 
are spanned by $H_{\bm{u}}$ and $E_{\Bal}^{R,I}$ that satisfy $(\bm{u}, \bm{t}) = (\Bal, \bm{t}) = 0 \FOR \forall H_{\bm{t}} \in L_H \cap L_G^c$. 
Therefore, $\Gh^\perp$ forms a subalgebra of $\Gg$. 
Since both $\Gh_C^\perp$ and $L_H \cap L_G^c$ are generated by the Cartan generators, $\Gh_C^\perp$ commutes with $L_H \cap L_G^c$. 
For $H_{\bm{t}} \in L_H \cap L_G^c$ and $E_{\Bal}^R, E_{\Bal}^I \in \Gh^\perp$, 
we have $\rbr{H_{\bm{t}}, E_{\Bal}^{R,I}} = \pm i (\Bal, \bm{t}) E_{\Bal}^{I,R} = 0$ from \refE{defgfp}, 
indicating that $L_H \cap L_G^c$ commutes with $\Gh^\perp$. 

Let $\sig$ be a representative element of $[\sig] \in \dis$. 
We next prove $\mr{Ad}(\sig)\Gh_C^\perp \subset \Gh^\perp$, 
where $\mr{Ad}(g): \Gg \to \Gg$ for $g \in G$ is defined by $\rbr{\mr{Ad}(g)}(X) := g X g^{-1} \FOR X \in \Gg$. 
For $H_{\bm{u}} \in \Gh_C^\perp$, 
we expand $\Ad(\sig) H_{\bm{u}}$ in terms of the Cartan canonical form as 
\ceqn{
\Ad(\sig) H_{\bm{u}} = H_{\bm{u}\pri} + \sum_{\Bal \in R_+} \br{ c_{\Bal} E_{\Bal}^R + d_{\Bal} E_{\Bal}^I}
}
where $c_{\Bal}$ and $d_{\Bal}$ are real numbers. 
Let $H_{\bm{t}}$ be an element of $L_H \cap L_G^c$. 
Since $\Ad(\sig)$ is an automorphism on $(L_H \cap L_G^c)/L_H^c$ from \refTH{theo2}, 
we have $\Ad(\sig^{-1})H_{\bm{t}} \in L_H \cap L_G^c$
and hence it can be written as $H_{\bm{t}} = \Ad(\sig)H_{\bm{t}\pri}$ for some $H_{\bm{t}\pri} \in L_H \cap L_G^c$. 
From \refE{defgcfp}, we have 
\peqn{
(\bm{t}, \bm{u}\pri) &=& \mr{Tr}\rbr{H_{\bm{t}} H_{\bm{u}\pri}}
 = \mr{Tr}\rbr{\Ad(\sig)H_{\bm{t}\pri} \Ad(\sig)H_{\bm{u}}} 
\nnco
&=& (\bm{t}\pri, \bm{u}) = 0
}
Hence we obtain $H_{\bm{u}\pri} \in \Gh_C^\perp$. 
Also, it follows from \refE{defgfp} that 
\peqn{
0 &=& \Ad(\sig)[H_{\bm{t}},  H_{\bm{u}}] 
\nnco
 &=& \rbr{H_{\bm{t}\pri},  H_{\bm{u}\pri} + \sum_{\Bal \in R_+} \br{ c_{\Bal} E_{\Bal}^R + d_{\Bal} E_{\Bal}^I}} 
\nnco
 &=& \sum_{\Bal \in R_+} i(\Bal, \bm{t}\pri) (c_{\Bal} E_{\Bal}^I - d_{\Bal} E_{\Bal}^R)
}
This gives $c_{\Bal} = d_{\Bal} = 0$ if $\Bal$ satisfies $(\Bal, \bm{t}\pri) \ne 0$ for some $\bm{t}\pri \in L_H \cap L_G^c$, 
resulting in $\Ad(\sig) H_{\bm{u}} \in \Gh^\perp$, 
which completes the proof of $\mr{Ad}(\sig)\Gh_C^\perp \subset \Gh^\perp$. 

Let $T_G$ be a maximum Abelian group of $G$. 
Let $H^\perp$ be the connected Lie group generated by $\Gh^\perp$ including the identity element. 
Since $\mr{Ad}(\sig)\Gh_C^\perp \subset \Gh^\perp$, 
$\Gh_C^\perp$ and $\Ad(\sig)\Gh_C^\perp$ are maximum Abelian subgroups of $H^\perp$. 
Since any two maximum Abelian subgroups are conjugate to each other \cite{Hall15, Brocker85}, 
there exists an element $h_\sig^\perp$ of $H^\perp$ such that $\Ad(h_\sigma^\perp)\rbr{\Ad(\sig)\Gh_C^\perp} = \Gh_C^\perp$. 
On the other hand, $\Ad(h_\sigma^\perp)$ acts on $L_H \cap L_G^c$ trivially from the commutativity between $\Gh^\perp$ and $L_H \cap L_G^c$. 
We therefore have $\Ad(h_\sigma^\perp) \Ad(\sig) L_H \cap L_G^c = L_H \cap L_G^c$. 
Since $\Gg_C$ is generated by $L_H \cap L_G^c$ and $\Gh_C^\perp$, 
we have 
\neqn{
h_\sigma^\perp \sig \in N_W := \tbr{g \in G | \mr{Ad}(g) X \in \Gg_C \FOR \forall X \in \Gg_C}
. \nnco
}
It is known that $T_G$ is a normal subgroup of $N_W$ 
and that the quotient group $N_W/T_G$ is isomorphic to $W_G$ \cite{Hall15, Brocker85}. 
We define $w_{\rbr{\sig}} \in W_G$ as the projection of $h_\sigma^\perp \sig \in N_W$ to $W_G \simeq N_W/T_G$. 
From \refTH{theo2}, 
the action of $(\rbr{H_{\bm{t}}}, \rbr{\sig}) \in \po$ on $H_{\bm{t}} + L_H^c \in \pt$ can be written as 
\peqn{
&&\lam_2^{(\rbr{H_{\bm{t}}},\rbr{\sig})}(H_{\bm{t}} + L_H^c)
\nnco
 &=& \lam_2^{(e, \rbr{\sig})}(H_{\bm{t}} + L_H^c) 
 = \Ad(\sig)(H_{\bm{t}}) + L_H^c 
\nnco
 &=& w_{[\sig]}(H_{\bm{t}}) + L_H^c
}
We note that the right-hand side does not depend on the choice of a representative element since $\Ad(\sig)$ acts on $L_H^c$ trivially. 
Thus we have 
\peqn{
\mathcal{G}_2 &\simeq& \tbr{\left. \lam_2^{(\rbr{H_{\bm{t}}},\rbr{\sig})} \right| (\rbr{H_{\bm{t}}}, \rbr{\sig}) \in \po}
\nnco
 &\simeq& \tbr{w_{[\sig]} \in W_G| [\sig] \in \dis}
}
Since the right-hand side is a subgroup of $W_G$, $\mathcal{G}_2$ is also a subgroup of $W_G$, 
which completes the proof of \refCO{cor5}. 


\bibliography{citation}
\end{document}